# Millimeter-wave spectroscopy of the $^{13}CH_3OD$ isotopic species of methyl alcohol

Li-Hong Xu[a], R.M. Lees[a]*, O. Zakharenko[b], H.S.P. Müller[b], F. Lewen[b], S. Schlemmer[b], K.M. Menten[c]

[a]*Centre for Laser, Atomic and Molecular Physics (CLAMS), Department of Physics, University of New Brunswick, 100 Tucker Park Road, Saint John, NB, Canada E2L 4L5*

[b]*I. Physikalisches Institut, Universität zu Köln, 50937 Köln, Germany*

[c]*Max-Planck-Institut für Radioastronomie, 53121 Bonn, Germany*

## Abstract

The dramatic increase in sensitivity, spectral coverage and resolution of radio astronomical facilities in recent years has opened new possibilities for observation of chemical differentiation and isotopic fractionation in protostellar sources to shed light on their spatial and temporal evolution. In warm interstellar environments, methanol is an abundant species, hence spectral data for its isotopic forms are of special interest. In the present work, the millimeter-wave spectrum of the $^{13}CH_3OD$ isotopologue has been investigated over the region from 150-510 GHz to provide a set of transition frequencies for potential astronomical application. The focus is on two types of prominent $^{13}CH_3OD$ spectral groupings, namely the *a*-type $^{q}R$-branch multiplets and the *b*-type *Q*-branches. Line positions are reported for the $^{q}R(J)$ clusters for $J$ = 3 to 10 for the $v_t$ = 0 and 1 torsional states, and for a number of $v_t$ = 0 and 1 $^{r}Q(J)$ or $^{p}Q(J)$ line series up to $J$ = 25. The frequencies have been fitted to a multi-parameter torsion-rotation Hamiltonian, and upper level excitation energies have been calculated from the resulting molecular constants.

## 1. Introduction

The dramatic advances in sensitivity, detector bandwidth and resolution of radio astronomical facilities in recent years have greatly enhanced the study of the intricate molecular environments in star-forming regions [1]. Numerous detailed astronomical surveys are highlighting the importance of the physics and chemistry of the protostellar stages, with observations of chemical differentiation and isotopic fractionation shedding new light on the spatial and temporal evolution and the physical conditions of the warm and dense environments of newly formed stars [1-7].

Since its discovery in the interstellar medium 50 years ago [8], methanol has been an important species in astronomical studies of interstellar clouds and star-forming regions, with its rich spectrum providing a valuable probe into the thermal and spatial structures of the sources [9]. As the high abundance of methanol renders numerous lines of the main $^{12}CH_3{}^{16}OH$ species optically thick in many sources, the C-13, O-18 and deuterated isotopologues have also been important in permitting observations deep into the source regions in order to obtain more accurate determinations of column densities (molecules per cm$^2$) and abundance profiles as well as information on astrochemical processes [1,2,4-7,9,10]. While the terrestrial and general cosmic abundance of deuterium is very low, much higher D/H ratios can be observed ranging up to several percent in warm protostellar regions with active chemistry [1,2,4-7,10,11]. Thus, the importance of deuterated isotopologues is significantly enhanced, providing motivation for the present study of $^{13}CH_3OD$.

The only previous spectroscopic data for $^{13}CH_3OD$ are from the investigation of the fundamental $J = 1 \leftarrow 0$ $a$-type $^{q}R_0$ transitions of numerous methanol isotopic species by Venkateswarlu et al. [12]. This study provided effective $B$ rotational constants for the $A$ and $E$ torsional symmetry species for the $v_t = 0$ to 2 torsional states. In the present work, we have extended observation and analysis to the $^{q}R_K(J)$ multiplets from $J = 3$ to 10 and $K = 0$ to 10 for the $v_t = 0$, 1 ground and first excited torsional states, and have also explored the $b$-type $^{r}Q(J)$ and $^{p}Q(J)$ $v_t = 0$ and $v_t = 1$ $Q$-branches lying in the spectral region from 150 to 510 GHz. Both types of spectral feature have characteristic "fingerprint" structures, giving a dataset with optimum possibility for astronomical detection. The data have been analyzed with a multi-parameter global fitting program, thereby providing upper-state excitation energies and a set of molecular constants that is compared to those for related isotopic species.

## 2. Experimental Details

Measurements over the 150 to 510 GHz millimeter-wave (MMW) region were carried out on the Cologne MMW spectrometer, details of which have been reported previously [13,14]. The MMW radiation was generated by a sequence of frequency multipliers (Virginia Diodes) driven by a MW synthesizer locked to an atomic clock. Schottky detectors were employed, with lock-in amplification in 2$f$ mode. The free-space absorption cell was a 5-m Pyrex tube, double-passed via a roof-top reflector, with a Teflon Brewster window at the entrance to the cell to separate the incident and reflected beams.

The $^{13}CH_3OD$ sample was produced by taking advantage of both the rapid D/H exchange at the hydroxyl group known for methanol and the strong methanol propensity for adsorption on cell surfaces. Prior to the start of the $^{13}CH_3OD$ study, an investigation of the $^{12}CH_3OD$ isotopologue had been ongoing in the 350-510 GHz region and served to precondition the cell walls. A sample of $^{13}CH_3OH$ was then mixed with an equal volume of $D_2O$ and the cell was filled with the vapour to a pressure of approximately 20 μbar. An initial sweep from 395-400 GHz showed a mixture of isotopologues, with strong $^{12}CH_3OD$ and $^{12}CH_3OH$ lines of comparable intensity together with weak $^{13}CH_3OD$ and $^{13}CH_3OH$ features down by about an order of magnitude. As we continued to higher frequencies with numerous sample refill cycles, however, the $^{12}CH_3OD$ was gradually replaced on the cell walls and pumped out, and the $^{13}CH_3OD$ lines became strongly dominant in the spectrum. We repeated the initial 395-400 GHz sweep a few days later, and the previous $^{12}CH_3OD$ and $^{12}CH_3OH$ features now had very low intensity.

For the majority of the multi-GHz sweeps, the step size between spectral points was set to 108 kHz. The JPL SMAP analysis program [15] was used for peak-fitting, but in some later cases the peak centres were interpolated directly from the point listings for convenience. In general, the accuracy of the line positions is estimated conservatively to be within ±0.05 MHz.

## 3. Spectral structure and notation

The only lines previously known for $^{13}CH_3OD$ were the $J_K = 1_0 \leftarrow 0_0$ $A/E$ doublets for the $v_t = 0$ to 2 torsional states [12], which provided effective $B$-values. Thus, to guide the initial search, values for the other major torsion-rotation parameters were estimated from those of the related $^{12}CH_3OH$, $^{12}CH_3OD$ and $^{13}CH_3OH$ species [16-18]. The spectral features could then be predicted employing a modified version of the BELGI program used for energy calculation and global fitting [16,19]. It was decided initially to target *a*-type *R*-branch multiplets and *b*-type *Q* branches, two classes of features that form distinctive line groupings in the spectrum [9] that would have the highest likelihood for successful astronomical detection. Both involve relatively close sets of transitions covering a broad range of excitation energy, providing distinctive fingerprints for identification as well as sensitivity to the physical conditions of an astronomical source.

In simplest form, the ground-state molecular torsion-rotation energies can be written as

$$E_{t\text{-}r}(v_t,TS,J,K) = E_t(v_t,TS,K) + B_{eff}J(J + 1) + (A - B)_{eff}K^2 + \text{higher-order terms} \quad (1)$$

where $E_t$ is the ($K$-dependent) torsional energy, $B_{eff}$ and $(A - B)_{eff}$ are effective rotational constants, $J$ is the overall rotational angular momentum with component $K$ along the molecular $a$-axis, $v_t$ is the torsional quantum number and $TS = A$ or $E$ denotes the torsional symmetry. For $|K| > 0$ for $E$ levels, a signed $K$ will be used with $K > 0$ and $K < 0$ levels forming different classes, while the $A$ levels with $K > 0$ are split by molecular asymmetry into $K$-doublets with the components distinguished as $A^+$ or $A^-$. The higher-order terms include centrifugal distortion, $J$-dependent asymmetry shifts and torsion-rotation interactions.

Transitions induced by the $\mu_a$ component of the molecular dipole moment have a $\Delta K = 0$ selection rule, whereas those induced by $\mu_b$ normally follow the rule $\Delta K = \pm 1$. Both types have $\Delta J = 0, \pm 1$ selection rules, denoted as $P$, $Q$ or $R$ for $\Delta J = -1$, 0 or +1, respectively. An initial $p$, $q$ or $r$ prefix is also used similarly to indicate $\Delta K$, giving the convenient notation $^qR_K(J)$ for a $(J + 1)_K \leftarrow J_K$ $a$-type $R$-branch transition and $^pQ_K(J)$ or $^rQ_K(J)$ for $J_{K-1} \leftarrow J_K$ or $J_{K+1} \leftarrow J_K$ $Q$-branch transitions.

For an $a$-type $R$-branch transition with $\Delta K = 0$ and $\Delta J = +1$, the basic frequency from Eq. (1) is independent of $K$, $TS$ and $v_t$ and is given simply as:

$$^qR(J) = 2B_{eff}(J + 1) \quad (2)$$

The higher-order terms then broaden this into a series of tight $v_t$-multiplets shifting down with increasing $v_t$, each containing $2(2J+1)$ transitions of differing $K$ and $TS$. Fig. 1 shows the $^{13}CH_3OD$ $^qR(8)$ $a$-type multiplets for $v_t = 0$ and $v_t = 1$, illustrating the close clustering in the spectrum. With $A$ and $E$ lines from $K = 0$ to 8, each $v_t$-multiplet contains 34 transitions with upper level excitation energies ranging from 66 to 255 cm$^{-1}$ for $v_t = 0$ and 242 to 457 cm$^{-1}$ for $v_t = 1$.

For a $K+1 \leftarrow K$ $Q$-branch, the $^rQ_K(J)$ members would all coincide in lowest order (as for a rigid symmetric-top molecule), but the higher-order terms introduce shifts strongly dependent on $J$. Thus, the typical $Q$ branch appears as a series of lines closely spaced near the origin and then gradually spreading as $J$ increases [7], as shown in Fig.2 for the $K = -2 \leftarrow -1$ $E$ $v_t = 0$ $Q$-branch. Three lines also appear in Fig. 2 from the $K = 2A^+ \leftarrow 1A^-$ $v_t = 0$ $Q$-branch that initially moves to low frequency but turns around at $J = 19$ and returns due to the differing $J$-dependence of the $K$- doubling for $K = 1$ and $K = 2$. The $K = -3 \leftarrow -2$ $E$ $v_t = 0$ $Q$-branch is illustrated in Fig. 3, initially shading to lower frequency, coming to a head at $Q$11, and then reversing due to competition among the different higher-order contributions as $J$ rises. (Note that the branch origin in the inset is at the upper limit of our spectral region where the MMW power is dropping off rapidly, so is rather difficult to see.)

## 4. Measurements

From the spectral predictions, we could select scan regions likely to contain significant $^{13}CH_3OD$ $R$-branch and/or $Q$-branch structures. Then, by comparing a template of the expected transition positions against the plotted spectra, we could search for those features via pattern recognition. An early success was the scan from 395-400 GHz in Fig.1, in which the clear line clusters corresponding to the $v_t = 0$ and 1 $^qR(8)$ $R$-branch multiplets are seen as well as several lines of the $K = 3 \leftarrow 2$ $E$ $v_t = 0$ $Q$-branch. The latter were within a few MHz of the predicted positions, hence we then proceeded to track the full $Q$-branch with short targeted individual 108 MHz scans for each line from the origin up to $J = 25$. As new assignments were made, they were then incorporated into the input dataset for the global fitting program in order to refine the molecular parameters and the predictions in an iterative cycle.

Proceeding in this way with a combination of broad multi-GHz scans plus targeted short scans for individual $Q$-branch lines and updating the predictions as we went, we were able to assign and measure all of the $A$ and $E$ $^qR_K(J)$ transitions from $J = 3$ to 10 with $K = 0$ to 10 for both $v_t = 0$ and 1, as well as 7 $Q$ branches for each of the $v_t = 0$ and 1 torsional states. Tables 1 and 2 list the $R$-branch transitions for $v_t = 0$ and $v_t = 1$, respectively, giving line positions, (o – c) residuals, upper level excitation energies for the transitions and relative intensity factors. Here the lines are ordered successively by $J$, torsional symmetry and then $K$, but a line list in increasing frequency order is also provided in the supplementary material. For the $Q$ branches, we tracked all lines within our observational range of 150-510 GHz up to a maximum $J$ of 25. The results are presented in Tables 3 and 4 for vt = 0 and vt = 1, respectively, ordered by torsional symmetry, $K$ and then $J$.

For the $v_t = 0$ $E$ levels from $K = 0$ to 2, as seen previously for $^{12}CH_3OD$ [17] and $^{12}CD_3OD$ [20], the well-known $K$-labeling difficulties [17,19-23] arose in attaching appropriate $K$ values to the computed energies to ensure that they met the intuitively reasonable criterion of following smooth trajectories with increasing $J$ with no level crossings or sudden jumps. The problem is illustrated here in Fig. 4 for the $J$-reduced energies given by $[E_{t\text{-}r} – 0.738J(J + 1)]$ where 0.736 $cm^{-1}$ is the effective $B$-value, showing several distinct anomalies in the $K$-labels assigned by the computer. (A similar problem occurs for $v_t = 1$ also where the computer has the $-2$ $E$ curve coming down and crossing the $-1$ $E$ curve between $J = 23$ and 24.) The issue is due primarily to the off-diagonal $\Delta K = \pm1$ and $\Delta K = \pm2$ asymmetry elements in the torsion-rotation matrix that couple the basis states to give $K$-mixed eigenfunctions. As these elements are strongly $J$-dependent, the mixing increases rapidly with $J$ and can lead to a change in $K$-character that triggers the computer $K$-labeling algorithm to suddenly switch between curves. In the spectrum, these alterations in $K$-character can modify the effective selection rules for affected transition series, resulting in significant changes in the intensity patterns. This is seen for example in Table 4 in the sharp reduction of the computed transition strengths for the $K$ = –2 ← –3 $E$ $v_t$=1 $Q$ branch as it evolves towards a "forbidden" $K$ = –1 ← –3 series at high $J$ due to the {–1/–2} $E$ mixing.

In the present work, we have manually rearranged the $K$-labelling, using Excel spreadsheet difference tables and charts to ensure smooth, non-crossing behavior for all substates as shown in Fig. 5, with correspondingly smooth $J$-progressions for all sub-band frequencies. The resulting table of adjusted energy levels is included in the supplementary material. However, a problem still remains for the spectral fitting in that the input data must be labelled in conformity with the computer's choice of $K$-values, otherwise large (o – c) residuals will result that can derail the fit [17]. The major issue here was the slow avoided crossing of the $K$ = 0 and –1 $E$ $v_t$ = 0 levels around $J$ = 17. This presented some difficulty to the energy calculation program, which output identical energies for the two $J$ = 17 levels (an impossibility) and interchanged the two states for $J$ > 17. This impacted our original measurements, because in following the computer $K$-labeling above $J$ = 17 for the "$K$ = –2 ← –1" $E$ $v_t$ = 0 $Q$-branch we were actually tracking lines of the $K$ = –2 ← 0 $E$ $Q$-branch that appears due to intensity borrowing induced by the asymmetry mixing. Subsequently, from the spreadsheet smoothing, we could deduce the missing $J$ = 17 $K$ = 0 $E$ energy, and then sort out and locate the further lines of the $K$ = –2 ← –1 and –2 ← 0 $E$ $v_t$ = 0 $Q$-branches that are given in Table 3.

Valuable confirmation of the $R$-branch and lower-$J$ $Q$-branch assignments was provided from closed combination loops of transitions, as illustrated in Fig. 6. The net change in energy going around a 4-sided loop should be zero, and, as shown in the figure caption, our observed loop defects are indeed close to zero to well within experimental uncertainty. For each of the reported $Q$ branches, this check was carried out for all possible loops containing $R$-branch partners within our observed range from $R$(3) to $R$(10) in order to confirm the assignments.

## 5. Spectral fitting and molecular parameters

In order to obtain molecular constants for $^{13}CH_3OD$, the $v_t$ = 0 and $v_t$ = 1 transition frequencies in Tables 1 to 4 plus the original $J = 1_0 \leftarrow 0_0$ lines from [12] were fitted employing a slightly modified version of the BELGI torsion-rotation program [16,19]. In lowest order, the basic form for the torsional energy of Eq.(1) in the RAM axis system used is

$$E_t = F\langle P_\gamma + \rho P_a\rangle^2 + V_3/2\, \langle 1 - \cos 3\gamma\rangle \qquad (3)$$

where $F$ is the internal rotation constant, $P_\gamma$ is the angular momentum of the methyl top, $V_3$ is the height of the torsional potential barrier and $\gamma$ is the angle of internal rotation. The parameter $\rho$ accounts for the fraction of overall axial angular momentum $K$ contributing to the top momentum, and acts as a scaling factor determining the period of oscillation of the torsional energies as a function of $K$. With incorporation of the molecular asymmetry term $(B – C)(P_b^2 – P_c^2)$ and the product of inertia term $D_{ab}(P_aP_b + P_bP_a)$ into Eq.(1), higher-order terms are then added to Eqs.(1) and (3) in power-series fashion as products of the torsional and rotational operators in successively higher order [16,19].

In total, 791 $^{a}R$-branch and *Q*-branch frequencies were measured, including 74 unresolved *A* *K*-doublet transitions. In the absence of a feature to treat the latter as single blended lines, both *K*-doublet components were included separately in the fit. Then, with the exclusion of the lines indicated by asterisks in Tables 1 to 4 that were affected by doublet broadening, partially blended with close neighbors or showed anomalous intensity or large (o - c) residuals, the final fit contained 779 transitions in all. The present observations were input with identical uncertainties, and the four early $1_0 \leftarrow 0_0$ lines were assigned uncertainties 10x greater. The final iteration of the fit incorporated 63 parameters and reproduced the data with an overall rms deviation of 0.101 MHz. The rms deviation for the $v_t = 0$ lines alone was 0.068 MHz, not far above the estimated experimental uncertainty of 0.05 MHz, while that for the weaker $v_t = 1$ lines was about twice as great at 0.125 MHz. The values obtained for the primary molecular constants are presented in Table 5, along with those of the related $^{12}CH_3OD$, $^{12}CH_3OH$ and $^{13}CH_3OH$ isotopologues. The $^{13}CH_3OD$ results are quite comparable to the $^{12}CH_3OD$ values, with small decreases in each term. A list of all the fitted $^{13}CH_3OD$ parameters is included in the supplementary material.

## 6. Conclusion

In this work, the MMW spectrum of the $^{13}CH_3OD$ isotopic species of methanol has been explored in the region from 150 to 510 GHz. The *R*3 to *R*10 multiplets of *a*-type $^{q}R$-branch transitions together with a total of 7 *b*-type $^{p}Q$ and $^{r}Q$ branches have been identified and measured for each of the $v_t = 0$ and $v_t = 1$ torsional states. These are strong and relatively compact features with characteristic spectral patterns that should serve as a good dataset for potential detection in sensitive astronomical surveys of important protostellar sources. Observations in star-forming regions of the $^{13}$CH3OD isotopologue together with its $^{13}CH_3OH$ and $^{12}CH_3OD$ relatives could shed light on the source 13C/12C and D/H ratios and give a further window into the astrochemical processes and the temporal and spatial evolution of the sources.

The observed transition frequencies have been analyzed with a multi-parameter torsion-rotation global fitting program that reproduces the data to close to measurement uncertainty. A valuable set of calculated ground-state level energies has also thereby been obtained from the computer output. This will serve as a base for future extension of the assignments to *b*-type *R*-branch and *P*-branch transitions and possibly higher torsional levels to expand the dataset and give a comprehensive spectral picture of the $^{13}CH_3OD$ ground state.

## Acknowledgments

R.M.L. and L.-H. Xu received financial support from the Natural Sciences and Engineering Research Council of Canada. R.M.L. expresses great appreciation for the warm hospitality and friendship extended to him and Li-Hong by Stephan and the whole Köln group during memorable visits. The work in Köln has been supported by the Deutsche Forschungsgemeinschaft (DFG) through the collaborative research grant SFB 956 (project ID 184018867), (sub-project B3), and through the Gerätezentrum "Cologne Center for Terahertz Spectroscopy" (project ID SCHL341/5-1).

**References**

[1] J.K. Jørgensen, A. Belloche, R.T. Garrod, Astrochemistry During the Formation of Stars, Annu. Rev. Astron. Astrophys. 58 (2020) 727-778, and references therein, https://doi.org/10.1146/annurev-astro-032620-021927.

[2] M.L. van Gelder, B. Tabone, L.Tychoniec, E.F. van Dishoeck, H. Beuther, A.C.A. Boogert, A. Caratti o Garatti, P.D. Klaassen, H. Linnartz, H.S.P. Müller, V. Taquet, Complex organic molecules in low-mass protostars on Solar System scales. I. Oxygen-bearing species**,** Astron Astrophys. 639 (2020) A87, https://doi.org/10.1051/0004-6361/202037758.

[3] A. Belloche, A. J. Maury, S. Maret, S. Anderl, A. Bacmann, Ph. André, S. Bontemps, S. Cabrit, C. Codella, M. Gaudel, F. Gueth, C. Lefèvre, B. Lefloch, L. Podio, L. Testi, Questioning the spatial origin of complex organic molecules in young protostars with the CALYPSO survey**,** Astron. Astrophys.635 (2020) A198, https://doi.org/10.1051/0004-6361/201937352.

[4] J.K. Jørgensen, H.S.P. Müller, H. Calcutt, A. Coutens, M.N. Drozdovskaya, K.I. Öberg, M.V. Persson, V. Taquet, E.F. van Dishoeck, S.F. Wampfler, The ALMA-PILS survey: isotopic composition of oxygen-containing complex organic molecules toward IRAS 16293-2422B**,** Astron. Astrophys. 620 (2018) A170, https://doi.org/10.1051/0004-6361/201731667.

[5] J.K. Jørgensen, M.H.D. van der Wiel, A. Coutens, J.M. Lykke, H.S.P. Müller, E.F. van Dishoeck, H. Calcutt, P. Bjerkeli, T.L. Bourke, M.N. Drozdovskaya, C. Favre, E.C. Fayolle, R.T. Garrod, S.K. Jacobsen, K.I. Öberg, M.V. Persson, S.F. Wampfler, The ALMA Protostellar Interferometric Line Survey (PILS). First results from an unbiased submillimeter wavelength line survey of the Class 0 protostellar binary IRAS 16293-2422 with ALMA, Astron. Astrophys. 595 (2016) A117, https://doi.org/10.1051/0004-6361/201628648.

[6] A. Belloche, H.S.P. Müller, R.T. Garrod, K.M. Menten, Exploring molecular complexity with ALMA (EMoCA): Deuterated complex organic molecules in Sagittarius B2(N2), Astron. Astrophys. 587 (2016) A91, https://doi.org/10.1051/0004-6361/201527268.

[7] H.S.P. Müller, A. Belloche, Li-Hong Xu, R.M. Lees, R.T. Garrod, A. Walters, J. van Wijngaarden, F. Lewen, S. Schlemmer, K.M. Menten, , Exploring molecular complexity with ALMA (EMoCA): Alkanethiols and alkanols in Sagittarius B2(N2), Astron. Astrophys. 587 (2016) A92, https://doi.org/10.1051/0004-6361/201527470.

[8] J.A. Ball, C.A. Gottlieb, A.E. Lilley, H.E. Radford, Detection of Methyl Alcohol in Sagittarius, Astrophys. J. Lett. 162 (1970) L203−L210, https://doi.org/10.1086/180654.

[9] S. Wang et al., , Herschel observations of EXtra-Ordinary Sources (HEXOS): Methanol as a probe of physical conditions in Orion KL, Astron. Astrophys. 527 (2011) A95, https://doi.org/10.1051/0004-6361/201015079.

[10].J.L. Neill, N.R. Crockett, E.A. Bergin, J.C. Pearson. Li-Hong Xu, Deuterated Molecules in Orion KL from Herschel/HIFI, Astrophys. J. 777 (2013) 85, https://doi.org/10.1088/0004-637X/777/2/85.

[11] B. Parise, C. Ceccarelli, A.G.G.M. Tielens, A. Castets, E. Caux, B. Lefloch, S.Maret, Testing grain surface chemistry: a survey of deuterated formaldehyde and methanol in low-mass class 0 protostars, Astron. Astrophys. 453 (2006) 949−958, https://doi.org/10.1051/0004-6361:20054476.

[12] P. Venkateswarlu, H.D. Edwards, W. Gordy, Methyl Alcohol. I. Microwave Spectrum, J. Chem. Phys. 23 (1955) 1195−1199, https://doi.org/10.1063/1.1742239.

[13] J.-B. Bossa, M.H. Ordu, H.S.P. Müller, F. Lewen, S. Schlemmer, , Laboratory spectroscopy of 1,2-propanediol at millimeter and submillimeter wavelengths, Astron. Astrophys. 570 (2014) A12, https://doi.org/10.1051/0004-6361/201424320.

[14] V.V. Ilyushin, O. Zakharenko, F. Lewen, S. Schlemmer, E.A. Alekseev, M. Pogrebnyak, R.M. Lees, Li-Hong Xu, A. Belloche, K.M. Menten, R.T. Garrod, H.S.P. Müller, Rotational spectrum of isotopic methyl mercaptan, $^{13}CH_3SH$, in the laboratory and towards Sagittarius B2(N2), Can. J. Phys. 98 (2020) 530−537, https://doi.org/10.1139/cjp-2019-0421.

[15] See http://spec.jpl.nasa.gov/ftp/pub/calpgm/SMAP/ for download of the SMAP spectral analysis program available from JPL.

[16] Li-Hong Xu, J. Fisher, R. M. Lees, H.Y. Shi, J.T. Hougen, J.C. Pearson, B.J. Drouin, G. A. Blake, R. Braakman, Torsion rotation global analysis of the first three torsional states ($v_t$ = 0, 1, 2) and terahertz database for methanol, J. Mol. Spectrosc. 251 (2008) 305−313, https://doi.org/10.1016/j.jms.2008.03.017.

[17] M.S. Walsh, Li-Hong Xu, R.M. Lees, I. Mukhopadhyay, G. Moruzzi, B.P. Winnewisser, S. Albert, Rebecca A.H. Butler, F.C. DeLucia, Millimeter-Wave Spectra and Global Torsion-Rotation Analysis for the $CH_3OD$ Isotopomer of Methanol, J. Mol. Spectrosc. 204 (2000) 60-71, https://doi.org/10.1006/jmsp.2000.8201.

[18] Li-Hong Xu, R.M. Lees, Yun Hao, H.S.P. Müller, C.P. Endres, F. Lewen, S. Schlemmer, K.M. Menten, Millimeter wave and terahertz spectra and global fit of torsion-rotation transitions in the ground, first and second excited torsional states of $^{13}CH_3OH$ methanol, J. Mol. Spectrosc. 303 (2014) 1-7, https://doi.org/10.1016/j.jms.2014.06.005.

[19] I. Kleiner, Asymmetric-top molecules containing one methyl-like internal rotor: Methods and codes for fitting and predicting spectra, J. Mol. Spectrosc. 260 (2010) 1−18, https://doi.org/10.1016/j.jms.2009.12.011.

[20] L.-H. Xu, H.S.P. Müller, F.F.S. van der Tak, S. Thorwirth, The millimeter-wave spectrum of perdeuterated methanol, $CD_3OD$, J. Mol. Spectrosc. 228 (2004) 220−229, https://doi.org/10.1016/j.jms.2004.04.015.

[21] J. Ortigoso, I. Kleiner, J.T. Hougen, The *K*-rotational labeling problem for eigenvectors from internal rotor calculations: Application to energy levels of acetaldehyde below the barrier, J. Chem. Phys. 110 (1999) 11688−10699, https://doi.org/10.1063/1.479115.

[22] R.D. Suenram, G.Yu. Golubiatnikov, I. I. Leonov, J.T. Hougen, J. Ortigoso, I. Kleiner, G. T. Fraser, Reinvestigation of the Microwave Spectrum of Acetamide, J. Mol. Spectrosc. 208 (2001) 188−193, https://doi.org/10.1006/jmsp.2001.8377.

[23] V. Ilyushin, A new scheme of *K*-labeling for torsion-rotation energy levels in low-barrier molecules, J. Mol. Spectrosc. 227 (2004) 140−150, https://doi.org/10.1016/j.jms.2004.05.013.

[24] R.M. Lees, A Note on $CH_3OD$ Spectra, Ground-State Energies and FIR Laser Assignments, Int. J. Infrared Millimeter Waves 21 (2000) 1787−1805, https://doi.org/10.1023/A:1006727701859.

**List of Tables**

**Table 1**

Millimeter-wave frequencies, upper level energies ($E_u$) and line strengths ($\mu^2 S$) for ${}^{q}R_K(J)$ *a*-type *R*-branch transitions in the $v_t = 0$ ground torsional state of $^{13}CH_3OD$

**Table 2**

Millimeter-wave frequencies, upper level energies ($E_u$) and line strengths ($\mu^2 S$) for ${}^{q}R_K(J)$ *a*-type *R*-branch transitions in the $v_t = 1$ first excited torsional state of $^{13}CH_3OD$

**Table 3**

Millimeter-wave frequencies, upper level energies ($E_u$) and line strengths ($\mu^2 S$) for ${}^{r}Q_K(J)$ and ${}^{p}Q_K(J)$ *b*-type *Q*-branch transitions in the $v_t = 0$ ground torsional state of $^{13}CH_3OD$

**Table 4**

Millimeter-wave frequencies, upper level energies ($E_u$) and line strengths ($\mu^2 S$)for ${}^{r}Q_K(J)$ and ${}^{p}Q_K(J)$ *b*-type *Q*-branch transitions in the $v_t = 1$ first excited torsional state of $^{13}CH_3OD$

**Table 5**

Principal fitted molecular parameters (in $cm^{-1}$) for $^{13}CH_3OD$ and comparison to related methanol isotopologues

**List of Figures**

**Fig. 1**.Section of the $^{13}CH_3OD$ MMW spectrum showing $^{q}R(8)$ *a*-type multiplets for $v_t = 0$ and $v_t = 1$.

**Fig. 2.** Section of the $^{13}CH_3OD$ MMW spectrum showing the origin of the $K = -2 \leftarrow -1$ *E* $v_t = 0$ *Q*-branch from $J = 2$ to 7, along with several lines from the $K = 2A^{+} \leftarrow 1A^{-}$ and $1A^{-} \leftarrow 0\ A^{+}$ $v_t = 0$ *Q*-branches.

**Fig. 3.** Section of the $^{13}CH_3OD$ MMW spectrum showing the origin region for the $K = -3 \leftarrow -2$ *E* $v_t = 0$ *Q*-branch from $J = 3$ to 14.

**Fig. 4**. *J*-reduced low-*K* $v_t = 0$ energy levels of *E* symmetry for $^{13}CH_3OD$, showing the original *K*-labeling assigned by the computer. Strong *K*-mixing at high *J* by asymmetry matrix elements leads to an apparent level crossing between the lower $K = 0E$ and -1*E* states around $J = 17$ and sudden jumps in *K*-labelling along the $K = 0$ and ±2 curves for $J > 25$.

**Fig. 5**. *J*-reduced low-*K* $v_t = 0$ energy levels of *E* symmetry for $^{13}CH_3OD$ with *K*-labeling empirically adjusted to remove the $K = 0E$ and -1*E* level crossing at $J = 17$ and give smooth variation along all substate energy curves, analogous to the pattern for $^{12}CH_3OD$ [24].

**Fig. 6**. Combination loop diagram confirming assignments for $K = 2A$ and 3*A* *R*-branch and *Q*-branch $v_t = 0$ transitions of $^{13}CH_3OD$. Transition frequencies (in MHz) are: $a = 488126.413$, $b = 398547.526$, $c = 486693.816$, $d = 399980.113$, $e = 485840.596$, $f = 405794.833$, $g = 486582.463$, and $h = 405052.950$. Combination defect for loop A (blue solid arrows) is $\delta_A = a + b - c - d = 0.010$ MHz. Defect for loop B (red dotted arrows) is $\delta_B = e + f - g - h = 0.016$ MHz.

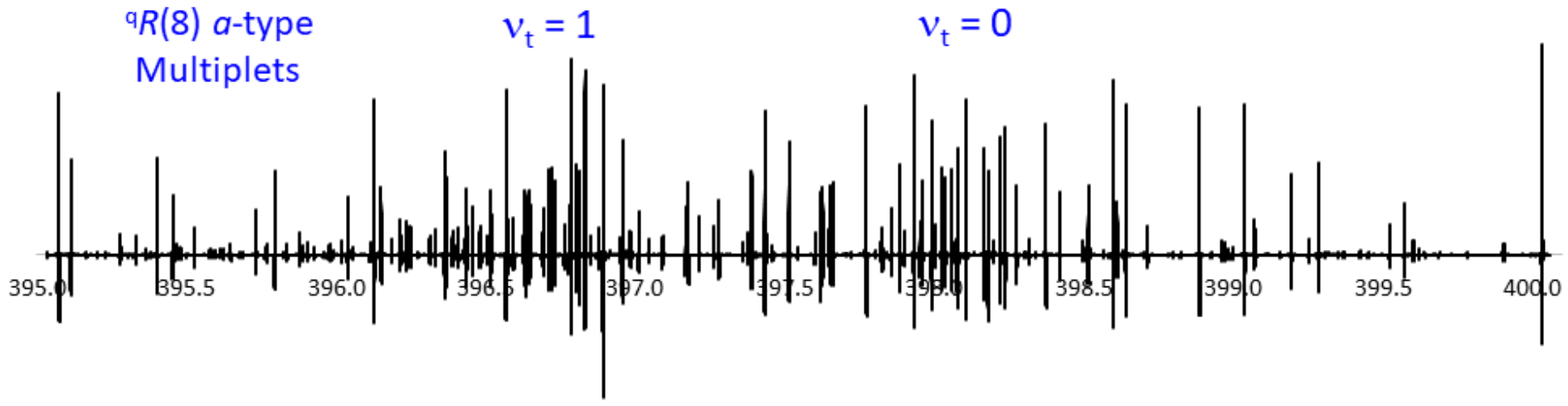

**Fig. 1**.Section of the $^{13}CH_3OD$ MMW spectrum showing $^{q}R(8)$ $a$-type multiplets for $v_t = 0$ and $v_t = 1$.

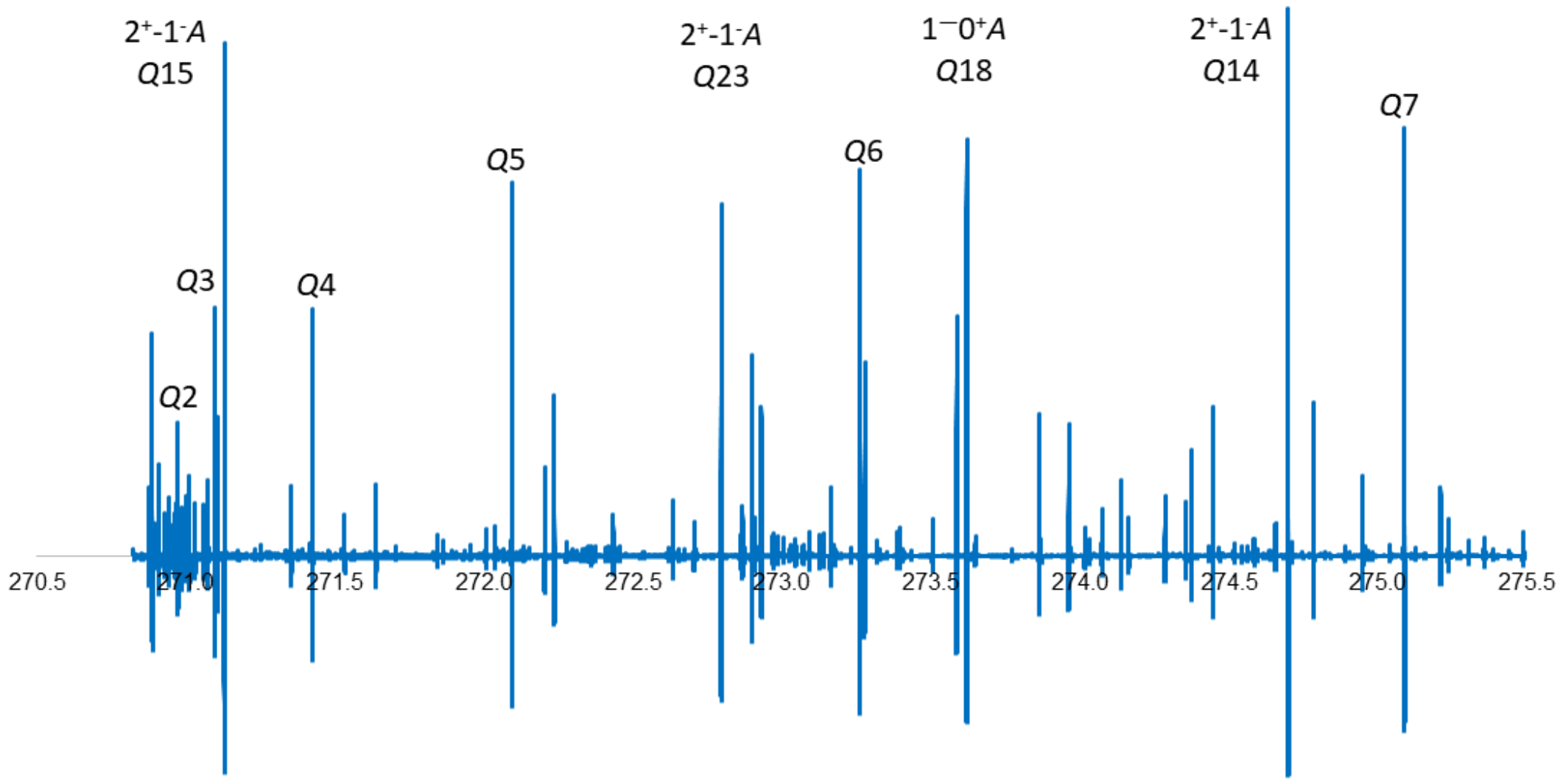

**Fig. 2.** Section of the $^{13}CH_3OD$ MMW spectrum showing the origin region of the $K = -2 \leftarrow -1$ $E$ $v_t = 0$ $Q$-branch from $J = 2$ to 7, along with several lines from the $K = 2A^+ \leftarrow 1A^-$ and $1A^- \leftarrow 0$ $A^+$ $v_t = 0$ $Q$-branches.

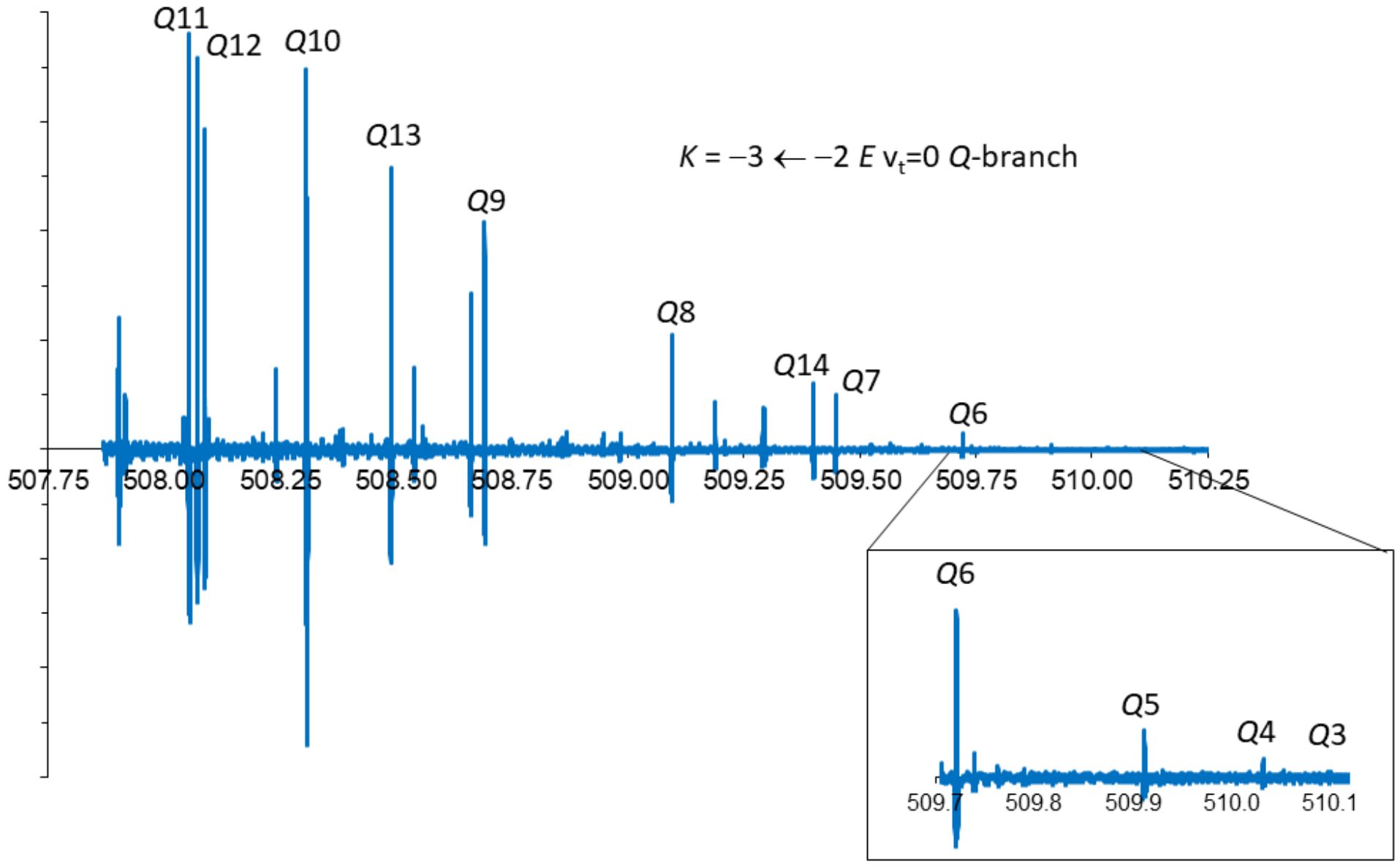

**Fig. 3.** Section of the $^{13}CH_3OD$ MMW spectrum showing the origin region for the $K = -3 \leftarrow -2$ $E$ $v_t = 0$ $Q$-branch from $J = 3$ to 14.

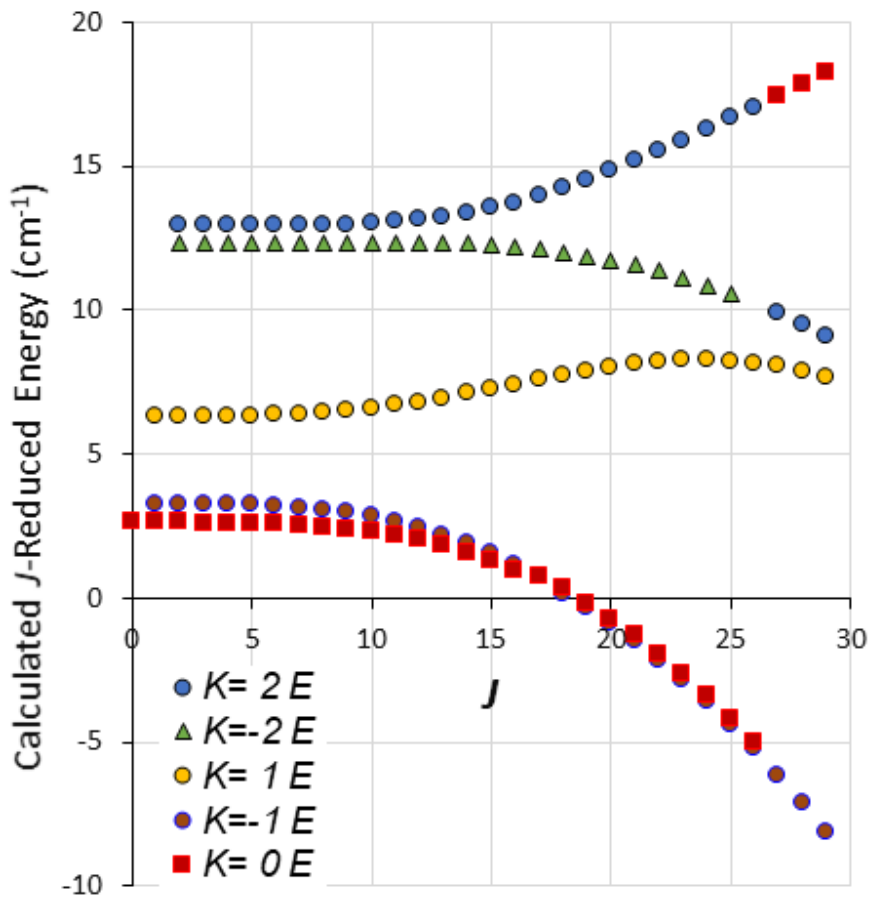

**Fig. 4**. *J*-reduced low-*K* $v_t = 0$ energy levels of *E* symmetry for $^{13}CH_3OD$, showing the original *K*-labeling assigned by the computer. Strong *K*-mixing at high *J* by asymmetry matrix elements leads to an apparent level crossing between the lower $K = 0E$ and $-1E$ states around $J = 17$ and sudden jumps in *K*-labelling along the $K = 0$ and $\pm 2$ curves for $J > 25$.

Chart Title

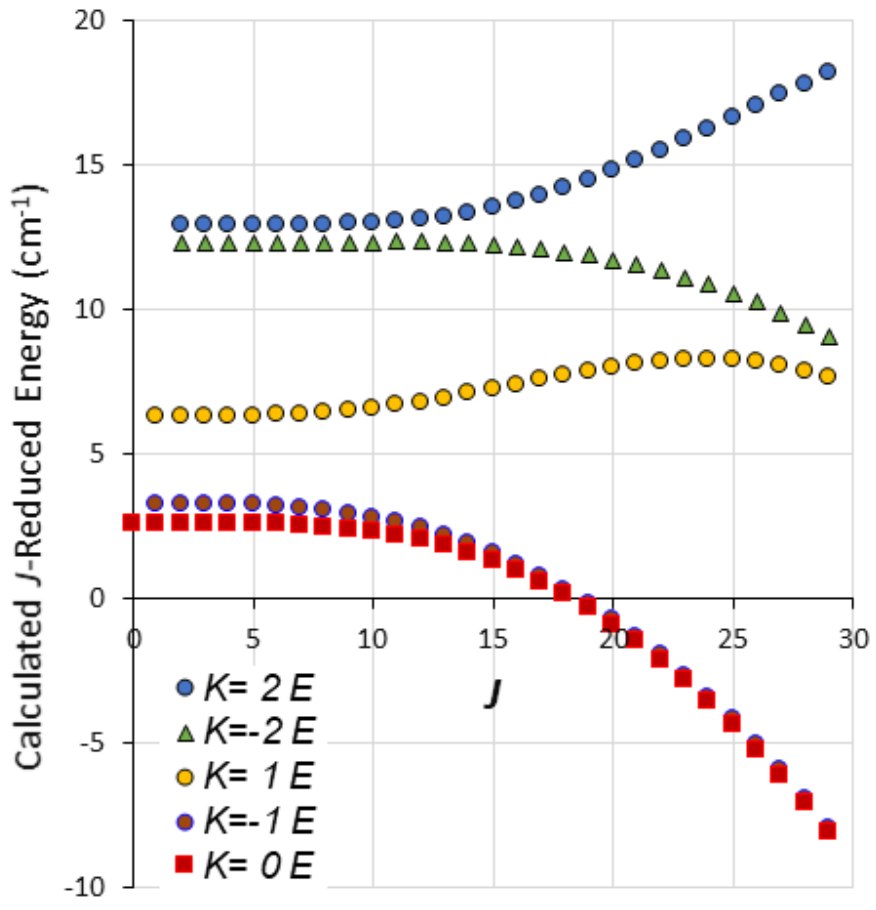

**Fig. 5**. *J*-reduced low-*K* $v_t = 0$ energy levels of *E* symmetry for $^{13}CH_3OD$ with *K*-labeling empirically adjusted to remove the $K = 0E$ and $-1E$ level crossing at $J = 17$ and give smooth variation along all substate energy curves, analogous to the pattern for $^{12}CH_3OD$ [24].

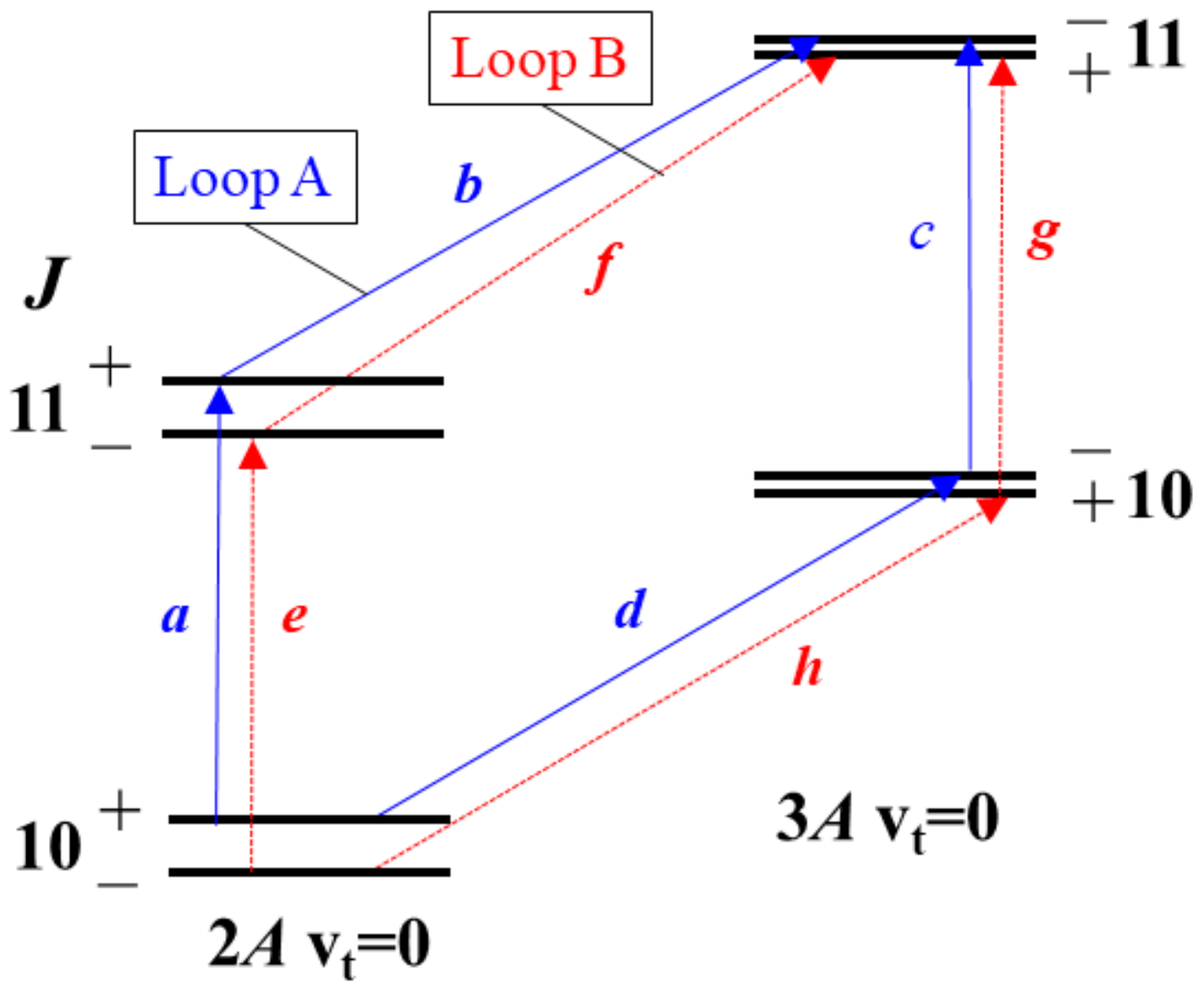

**Fig. 6**. Combination loop diagram confirming assignments for $K$ = 2$A$ and 3$A$ $R$-branch and $Q$-branch $v_t$ = 0 transitions of $^{13}CH_3OD$. Transition frequencies (in MHz) are: $a$ = 488126.413, $b$ = 398547.526, $c$ = 486693.816, $d$ = 399980.113, $e$ = 485840.596, $f$ = 405794.833, $g$ = 486582.463, and $h$ = 405052.950. Combination defect for loop A (blue solid arrows) is $\delta_A = a + b - c - d$ = 0.010 MHz. Defect for loop B (red dotted arrows) is $\delta_B = e + f - g - h$ = 0.016 MHz.

**Table 1**

Millimeter-wave frequencies, upper level energies ($E_u$) and line strengths ($\mu^2 S$) for $^{q}R_K(J)$ *a*-type *R*-branch transitions in the $v_t = 0$ ground torsional state of $^{13}CH_3OD$

| *J'* - *J''* | *K* | *TS* | $\nu_{obs}$ (MHz) | (o-c)[a] | $E_u$ (cm$^{-1}$)[a] | $\mu^2$S[c] | *J'* - *J''* | *K* | *TS* | $\nu_{obs}$ (MHz) | (o-c)[a] | $E_u$ (cm$^{-1}$)[b] | $\mu^2$S[c] |
|---|---|---|---|---|---|---|---|---|---|---|---|---|---|
| 1 - 0 | 0 | *A*+ | 44241.43[d] | 0.058 | 1.4757 | 0.68 | 7 - 6 | 0 | *A*+ | 309036.218 | 0.032 | 41.2714 | 4.79 |
| 1 - 0 | 0 | *E* | 44228.26[d] | 0.045 | 4.0772 | 0.64 | 7 - 6 | 1 | *A*+ | 304933.513 | 0.022 | 45.1596 | 4.68 |
| 4 - 3 | 0 | *A*+ | 176848.386 | 0.019 | 14.7515 | 2.74 | 7 - 6 | 1 | *A*− | 313977.907 | 0.030 | 46.3676 | 4.67 |
| 4 - 3 | 1 | *A*+ | 174337.152 | 0.019 | 18.9999 | 2.56 | 7 - 6 | 2 | *A*+ | 310068.398 | 0.019 | 56.5624 | 4.33 |
| 4 - 3 | 1 | *A*− | 179514.003 | 0.032 | 19.4317 | 2.55 | 7 - 6 | 2 | *A*− | 309473.816 | 0.025 | 56.5177 | 4.33 |
| 4 - 3 | 2 | *A*+ | 177028.294 | 0.012 | 29.9744 | 2.02 | 7 - 6 | 3 | *A*+ | 309660.927 | 0.009 | 69.9871 | 3.84 |
| 4 - 3 | 2 | *A*− | 176921.667 | 0.012 | 29.9690 | 2.02 | 7 - 6 | 3 | *A*− | 309671.888 | 0.011 | 69.9877 | 3.84 |
| 4 - 3 | 3 | *A*+ | 176949.01 | 0.049* | 43.4263 | 1.18 | 7 - 6 | 4 | *A*+ | 309658.654 | -0.034 | 88.4201 | 3.21 |
| 4 - 3 | 3 | *A*− | 176949.43 | -0.040* | 43.4263 | 1.18 | 7 - 6 | 4 | *A*− | 309658.654 | 0.048 | 88.4201 | 3.21 |
| 4 - 3 | 0 | *E* | 176745.180 | 0.042 | 17.3465 | 2.56 | 7 - 6 | 5 | *A*± | 309612.249 | 0.014 | 115.5043 | 2.33 |
| 4 - 3 | 1 | *E* | 177173.969 | 0.020 | 21.0621 | 2.52 | 7 - 6 | 6 | *A*± | 309537.363 | -0.047* | 150.0557 | 0.92 |
| 4 - 3 | 2 | *E* | 177008.316 | 0.001 | 27.6732 | 2.03 | 7 - 6 | 0 | *E* | 308663.091 | 0.036 | 43.8399 | 4.46 |
| 4 - 3 | 3 | *E* | 176976.207 | 0.002 | 41.2089 | 1.19 | 7 - 6 | 1 | *E* | 310771.040 | 0.090 | 47.6983 | 4.61 |
| 4 - 3 | −1 | *E* | 176696.221 | -0.015 | 18.0005 | 2.54 | 7 - 6 | 2 | *E* | 309944.848 | 0.007 | 54.2528 | 4.35 |
| 4 - 3 | −2 | *E* | 177024.867 | 0.018 | 27.0542 | 2.04 | 7 - 6 | 3 | *E* | 309758.081 | 0.004 | 67.7765 | 3.90 |
| 4 - 3 | −3 | *E* | 176956.259 | 0.007 | 44.0671 | 1.18 | 7 - 6 | 4 | *E* | 309648.718 | -0.005 | 90.1485 | 3.20 |
| 5 - 4 | 0 | *A*+ | 220972.803 | 0.019 | 22.1223 | 3.42 | 7 - 6 | 5 | *E* | 309562.992 | 0.008 | 118.1565 | 2.30 |
| 5 - 4 | 1 | *A*+ | 217890.551 | 0.010 | 26.2680 | 3.27 | 7 - 6 | 6 | *E* | 309519.739 | -0.027 | 148.8213 | 1.25 |
| 5 - 4 | 1 | *A*− | 224359.449 | 0.019 | 26.9155 | 3.27 | 7 - 6 | −1 | *E* | 308131.389 | 0.008 | 44.4603 | 4.62 |
| 5 - 4 | 2 | *A*+ | 221338.029 | -0.013 | 37.3574 | 2.83 | 7 - 6 | −2 | *E* | 309960.163 | 0.022 | 53.6363 | 4.38 |
| 5 - 4 | 2 | *A*− | 221124.955 | -0.001 | 37.3449 | 2.83 | 7 - 6 | −3 | *E* | 309687.065 | 0.012 | 70.6297 | 3.86 |
| 5 - 4 | 3 | *A*+ | 221186.650 | 0.028 | 50.8043 | 2.15 | 7 - 6 | −4 | *E* | 309597.045 | -0.009 | 91.4177 | 3.16 |
| 5 - 4 | 3 | *A*− | 221188.444 | -0.006 | 50.8044 | 2.15 | 7 - 6 | −5 | *E* | 309585.078 | 0.010 | 115.5173 | 2.32 |
| 5 - 4 | 4 | *A*+ | 221185.629 | -0.007 | 69.2375 | 1.22 | 7 - 6 | −6 | *E* | 309567.818 | -0.010 | 147.0827 | 1.26 |
| 5 - 4 | 4 | *A*− | 221185.629 | -0.002 | 69.2375 | 1.22 | 8 - 7 | 0 | *A*+ | 352953.588 | 0.041 | 53.0447 | 5.48 |
| 5 - 4 | 0 | *E* | 220805.854 | 0.041 | 24.7118 | 3.20 | 8 - 7 | 1 | *A*+ | 348416.167 | -0.008 | 56.7815 | 5.37 |
| 5 - 4 | 1 | *E* | 221615.720 | 0.023 | 28.4544 | 3.22 | 8 - 7 | 1 | *A*− | 358740.668 | 0.033 | 58.3339 | 5.37 |
| 5 - 4 | 2 | *E* | 221291.468 | 0.013 | 35.0547 | 2.84 | 8 - 7 | 2 | *A*+ | 354501.246 | -0.001 | 68.3873 | 5.05 |
| 5 - 4 | 3 | *E* | 221229.939 | -0.012 | 48.5883 | 2.18 | 8 - 7 | 2 | *A*− | 353612.161 | 0.003 | 68.3129 | 5.05 |
| 5 - 4 | 4 | *E* | 221179.371 | 0.011 | 70.9665 | 1.22 | 8 - 7 | 3 | *A*+ | 353896.360 | 0.047* | 81.7918 | 4.62 |
| 5 - 4 | −1 | *E* | 220649.514 | -0.012 | 25.3605 | 3.25 | 8 - 7 | 3 | *A*− | 353918.231 | 0.019 | 81.7931 | 4.62 |
| 5 - 4 | −2 | *E* | 221318.654 | 0.015 | 34.4366 | 2.86 | 8 - 7 | 4 | *A*+ | 353894.679 | -0.126* | 100.2248 | 4.08 |
| 5 - 4 | −3 | *E* | 221197.982 | 0.016 | 51.4454 | 2.16 | 8 - 7 | 4 | *A*− | 353894.679 | 0.100* | 100.2248 | 4.08 |
| 5 - 4 | −4 | *E* | 221151.955 | -0.019 | 72.2386 | 1.21 | 8 - 7 | 5 | *A*± | 353835.193 | 0.009 | 127.3070 | 3.32 |
| 6 - 5 | 0 | *A*+ | 265039.112 | 0.027 | 30.9631 | 4.11 | 8 - 7 | 6 | *A*± | 353741.895 | 0.003 | 161.8553 | 2.31 |
| 6 - 5 | 1 | *A*+ | 261423.842 | -0.002 | 34.9881 | 3.98 | 8 - 7 | 7 | *A*± | 353652.671 | -0.090 | 199.7406 | 1.25 |
| 6 - 5 | 1 | *A*− | 269182.401 | 0.009 | 35.8945 | 3.97 | 8 - 7 | 0 | *E* | 352431.127 | 0.037 | 55.5957 | 5.08 |
| 6 - 5 | 2 | *A*+ | 265682.610 | 0.009 | 46.2196 | 3.59 | 8 - 7 | 1 | *E* | 355479.651 | 0.018 | 59.5559 | 5.29 |
| 6 - 5 | 2 | *A*− | 265310.223 | 0.025 | 46.1947 | 3.59 | 8 - 7 | 2 | *E* | 354347.030 | 0.022 | 66.0725 | 5.07 |
| 6 - 5 | 3 | *A*+ | 265424.121 | 0.012 | 59.6579 | 3.03 | 8 - 7 | 3 | *E* | 354035.028 | 0.012 | 79.5859 | 4.69 |
| 6 - 5 | 3 | *A*− | 265428.980 | -0.002 | 59.6581 | 3.03 | 8 - 7 | 4 | *E* | 353882.510 | -0.024 | 101.9528 | 4.07 |
| 6 - 5 | 4 | *A*+ | 265422.282 | -0.007 | 78.0910 | 2.27 | 8 - 7 | 5 | *E* | 353771.717 | -0.043 | 129.9571 | 3.27 |
| 6 - 5 | 4 | *A*− | 265422.282 | 0.018 | 78.0910 | 2.27 | 8 - 7 | 6 | *E* | 353718.461 | -0.055 | 160.6201 | 2.35 |
| 6 - 5 | 5 | *A*± | 265386.698 | 0.000 | 105.1768 | 1.25 | 8 - 7 | 7 | *E* | 353710.436 | 0.034* | 196.9174 | 1.27 |
| 6 - 5 | 0 | *E* | 264783.542 | 0.055 | 33.5440 | 3.84 | 8 - 7 | −1 | *E* | 351646.487 | -0.006 | 56.1900 | 5.28 |
| 6 - 5 | 1 | *E* | 266146.936 | 0.040 | 37.3321 | 3.92 | 8 - 7 | −2 | *E* | 354288.306 | 0.002 | 65.4541 | 5.10 |
| 6 - 5 | 2 | *E* | 265599.554 | 0.009 | 43.9141 | 3.61 | 8 - 7 | −3 | *E* | 353935.142 | -0.008 | 82.4357 | 4.64 |
| 6 - 5 | 3 | *E* | 265490.166 | 0.018 | 57.4441 | 3.07 | 8 - 7 | −4 | *E* | 353813.979 | -0.012 | 103.2196 | 4.02 |
| 6 - 5 | 4 | *E* | 265414.303 | 0.003 | 79.8197 | 2.26 | 8 - 7 | −5 | *E* | 353798.492 | -0.022 | 127.3188 | 3.30 |
| 6 - 5 | 5 | *E* | 265349.107 | -0.020 | 107.8306 | 1.23 | 8 - 7 | −6 | *E* | 353779.339 | 0.006 | 158.8835 | 2.38 |
| 6 - 5 | −1 | *E* | 264465.447 | 0.000 | 34.1821 | 3.94 | 8 - 7 | −7 | *E* | 353707.794 | -0.058* | 199.0790 | 1.27 |
| 6 - 5 | −2 | *E* | 265632.138 | 0.016 | 43.2972 | 3.63 | 9 - 8 | 0 | *A*+ | 396781.475 | 0.024 | 66.2799 | 6.16 |
| 6 - 5 | −3 | *E* | 265441.453 | 0.006 | 60.2996 | 3.04 | 9 - 8 | 1 | *A*+ | 391868.858 | -0.011 | 69.8529 | 6.07 |
| 6 - 5 | −4 | *E* | 265376.197 | 0.001 | 81.0906 | 2.23 | 9 - 8 | 1 | *A*− | 403465.121 | 0.018 | 71.7921 | 6.06 |
| 6 - 5 | −5 | *E* | 265367.070 | -0.005 | 105.1907 | 1.24 | 9 - 8 | 2 | *A*+ | 398986.256 | 0.018 | 81.6960 | 5.76 |

Table 1 (*continued*)

| *J*′ - *J*″ | *K* | *TS* | $\nu_{obs}$ (MHz) | (o-c)[a] | $E_u$ ($cm^{-1}$)[a] | $\mu^2S$[c] | *J*′ - *J*″ | *K* | *TS* | $\nu_{obs}$ (MHz) | (o-c)[a] | $E_u$ ($cm^{-1}$)[b] | $\mu^2S$[c] |
|---|---|---|---|---|---|---|---|---|---|---|---|---|---|
| 9 - 8 | 2 | *A*– | 397721.769 | 0.020 | 81.5795 | 5.76 | 10 - 9 | 7 | *E* | 442086.589 | 0.001 | 224.9364 | 3.46 |
| 9 - 8 | 3 | *A*+ | 398129.265 | -0.021* | 95.0719 | 5.38 | 10 - 9 | 8 | *E* | 442026.508 | -0.026 | 270.0279 | 2.44 |
| 9 - 8 | 3 | *A*– | 398169.406 | 0.015 | 95.0746 | 5.38 | 10 - 9 | 9 | *E* | 441888.190 | -0.294 | 322.0368 | 1.27 |
| 9 - 8 | 4 | *A*+ | 398130.449 | -0.181* | 113.5050 | 4.91 | 10 - 9 | –1 | *E* | 438282.382 | 0.011 | 83.9860 | 6.53 |
| 9 - 8 | 4 | *A*– | 398130.449 | 0.360* | 113.5050 | 4.91 | 10 - 9 | –2 | *E* | 442824.650 | 0.005 | 93.5207 | 6.51 |
| 9 - 8 | 5 | *A*+ | 398055.143 | -0.021 | 140.5847 | 4.23 | 10 - 9 | –3 | *E* | 442440.387 | -0.011 | 110.4760 | 6.15 |
| 9 - 8 | 5 | *A*– | 398055.143 | -0.024 | 140.5847 | 4.23 | 10 - 9 | –4 | *E* | 442233.883 | -0.012 | 131.2477 | 5.63 |
| 9 - 8 | 6 | *A*± | 397940.236 | -0.068 | 175.1291 | 3.36 | 10 - 9 | –5 | *E* | 442209.165 | -0.004 | 155.3454 | 5.07 |
| 9 - 8 | 7 | *A*± | 397833.665 | -0.101 | 213.0110 | 2.37 | 10 - 9 | –6 | *E* | 442186.696 | -0.013 | 186.9087 | 4.36 |
| 9 - 8 | 8 | *A*± | 397805.194 | 0.094 | 254.7112 | 0.00 | 10 - 9 | –7 | *E* | 442087.870 | -0.036 | 227.0980 | 3.45 |
| 9 - 8 | 0 | *E* | 396076.694 | 0.041 | 68.8074 | 5.68 | 10 - 9 | –8 | *E* | 441941.337 | -0.271* | 272.2938 | 2.40 |
| 9 - 8 | 1 | *E* | 400252.160 | 0.007 | 72.9068 | 5.98 | 10 - 9 | –9 | *E* | 441853.781 | 0.131* | 320.2916 | 1.27 |
| 9 - 8 | 2 | *E* | 398835.625 | -0.006 | 79.3763 | 5.77 | 11 - 10 | 0 | *A*+ | 484136.221 | 0.003 | 97.1228 | 7.53 |
| 9 - 8 | 3 | *E* | 398322.238 | -0.004 | 92.8725 | 5.45 | 11 - 10 | 1 | *A*+ | 478673.705 | 0.000 | 100.3394 | 7.46 |
| 9 - 8 | 4 | *E* | 398115.642 | 0.004 | 115.2325 | 4.90 | 11 - 10 | 1 | *A*– | 492775.053 | 0.015 | 103.1778 | 7.43 |
| 9 - 8 | 5 | *E* | 397974.683 | -0.046* | 143.2321 | 4.17 | 11 - 10 | 2 | *A*+ | 488126.413 | 0.020 | 112.7726 | 7.17 |
| 9 - 8 | 6 | *E* | 397909.896 | -0.031 | 173.8929 | 3.36 | 11 - 10 | 2 | *A*– | 485840.596 | 0.026 | 112.5222 | 7.17 |
| 9 - 8 | 7 | *E* | 397902.457 | 0.104* | 210.1900 | 2.41 | 11 - 10 | 3 | *A*+ | 486582.463 | 0.015 | 126.0581 | 6.85 |
| 9 - 8 | 8 | *E* | 397847.838 | -0.029 | 255.2835 | 1.28 | 11 - 10 | 3 | *A*– | 486693.816 | 0.011 | 126.0667 | 6.85 |
| 9 - 8 | –1 | *E* | 395022.707 | 0.012 | 69.3665 | 5.92 | 11 - 10 | 4 | *A*+ | 486601.503 | -0.008 | 144.4920 | 6.49 |
| 9 - 8 | –2 | *E* | 398589.637 | 0.008 | 78.7496 | 5.81 | 11 - 10 | 4 | *A*– | 486599.169 | -0.007 | 144.4919 | 6.49 |
| 9 - 8 | –3 | *E* | 398186.127 | 0.002 | 95.7177 | 5.40 | 11 - 10 | 5 | *A*+ | 486484.643 | -0.036 | 171.5647 | 5.94 |
| 9 - 8 | –4 | *E* | 398026.429 | -0.025 | 116.4964 | 4.84 | 11 - 10 | 5 | *A*– | 486484.643 | -0.054 | 171.5647 | 5.94 |
| 9 - 8 | –5 | *E* | 398006.753 | -0.010 | 140.5949 | 4.21 | 11 - 10 | 6 | *A*± | 486315.757 | -0.098 | 206.0988 | 5.20 |
| 9 - 8 | –6 | *E* | 397985.845 | 0.008 | 172.1589 | 3.40 | 11 - 10 | 7 | *A*± | 486167.551 | -0.123 | 243.9715 | 3.35 |
| 9 - 8 | –7 | *E* | 397901.270 | -0.134* | 212.3515 | 2.40 | 11 - 10 | 8 | *A*± | 486130.356 | 0.157 | 285.6694 | 3.08 |
| 9 - 8 | –8 | *E* | 397775.204 | -0.778* | 257.5522 | 1.26 | 11 - 10 | 9 | *A*± | 486100.257 | 0.193 | 335.3014 | 2.47 |
| 10 - 9 | 0 | *A*+ | 440511.360 | 0.015 | 80.9738 | 6.85 | 11 - 10 | 10 | *A*± | 485975.555 | -0.274* | 393.2796 | |
| 10 - 9 | 1 | *A*+ | 435288.842 | -0.004 | 84.3725 | 6.76 | 11 - 10 | 0 | *E* | 482973.845 | 0.011 | 99.5809 | 6.79 |
| 10 - 9 | 1 | *A*– | 448145.350 | 0.000 | 86.7406 | 6.75 | 11 - 10 | 1 | *E* | 489859.465 | 0.009 | 104.0923 | 7.34 |
| 10 - 9 | 2 | *A*+ | 443527.191 | -0.007 | 96.4905 | 6.47 | 11 - 10 | 2 | *E* | 488225.039 | -0.033 | 110.4536 | 7.15 |
| 10 - 9 | 2 | *A*– | 441799.058 | 0.013 | 96.3163 | 6.47 | 11 - 10 | 3 | *E* | 486932.609 | -0.004 | 123.8791 | 6.94 |
| 10 - 9 | 3 | *A*+ | 442358.539 | 0.003 | 109.8274 | 6.12 | 11 - 10 | 4 | *E* | 486579.263 | -0.050 | 146.2181 | 6.47 |
| 10 - 9 | 3 | *A*– | 442427.193 | 0.003 | 109.8324 | 6.12 | 11 - 10 | 5 | *E* | 486360.230 | -0.096 | 174.2045 | 5.85 |
| 10 - 9 | 4 | *A*+ | 442366.235 | 0.056* | 128.2607 | 5.71 | 11 - 10 | 6 | *E* | 486266.814 | -0.263* | 204.8597 | 5.19 |
| 10 - 9 | 4 | *A*– | 442364.930 | -0.080* | 128.2607 | 5.71 | 11 - 10 | 7 | *E* | 486262.503 | 0.254 | 241.1564 | 4.44 |
| 10 - 9 | 5 | *A*+ | 442271.778 | -0.015 | 155.3373 | 5.10 | 11 - 10 | 8 | *E* | 486196.699 | -0.069 | 286.2457 | 3.51 |
| 10 - 9 | 5 | *A*– | 442271.778 | -0.022 | 155.3373 | 5.10 | 11 - 10 | 9 | *E* | 486039.420 | -0.322* | 338.2493 | 2.43 |
| 10 - 9 | 6 | *A*± | 442131.790 | -0.091 | 189.8771 | 4.30 | 11 - 10 | 10 | *E* | 485872.679 | 0.170 | 393.3271 | 1.27 |
| 10 - 9 | 7 | *A*± | 442005.649 | -0.089 | 227.7547 | 3.36 | 11 - 10 | –1 | *E* | 481453.390 | 0.020 | 100.0456 | 7.09 |
| 10 - 9 | 8 | *A*± | 441972.957 | 0.126 | 269.4539 | 2.27 | 11 - 10 | –2 | *E* | 486950.772 | 0.010 | 109.7636 | 7.19 |
| 10 - 9 | 9 | *A*± | 441942.577 | 0.206* | 319.0868 | 1.29 | 11 - 10 | –3 | *E* | 486698.423 | -0.018 | 126.7105 | 6.88 |
| 10 - 9 | 0 | *E* | 439592.148 | 0.028 | 83.4706 | 6.26 | 11 - 10 | –4 | *E* | 486435.763 | -0.012 | 147.4735 | 6.40 |
| 10 - 9 | 1 | *E* | 445057.491 | 0.012 | 87.7524 | 6.66 | 11 - 10 | –5 | *E* | 486405.091 | 0.003 | 171.5701 | 5.90 |
| 10 - 9 | 2 | *E* | 443449.221 | 0.003 | 94.1681 | 6.46 | 11 - 10 | –6 | *E* | 486381.322 | 0.009 | 203.1326 | 5.26 |
| 10 - 9 | 3 | *E* | 442621.029 | 0.005 | 107.6367 | 6.20 | 11 - 10 | –7 | *E* | 486266.814 | 0.247* | 243.3181 | 4.42 |
| 10 - 9 | 4 | *E* | 442347.909 | -0.023 | 129.9876 | 5.70 | 11 - 10 | –8 | *E* | 486096.990 | -0.220 | 288.5083 | 3.45 |
| 10 - 9 | 5 | *E* | 442171.101 | -0.060 | 157.9813 | 5.03 | 11 - 10 | –9 | *E* | 485995.511 | 0.217 | 336.5027 | 2.43 |
| 10 - 9 | 6 | *E* | 442093.022 | -0.062 | 188.6396 | 4.30 | 11 - 10 | –10 | *E* | 485975.555 | 0.702* | 390.7932 | 1.29 |

[a] Observed minus calculated line frequencies in MHz. Asterisks indicate blended or barely resolved lines that were excluded from the fit.

[b] Calculated upper level energies in $cm^{-1}$. The zero reference energy is taken as the $J = K = 0$ *A*+ $v_t = 0$ level, calculated to lie 104.8883 $cm^{-1}$ above the bottom of the torsional barrier.

[c] *S* is the transition strength. Relative line intensities are given by the product of $\mu^2 S$ and the appropriate Boltzmann factor. The dipole moment components were taken as $\mu_a = 0.836$ and $\mu_b = -1.439$ Debye.

[d] Ref. [12].

## Table 2

Millimeter-wave frequencies, upper level energies ($E_u$) and line strengths ($\mu^2 S$) for ${}^{q}R_K(J)$ *a*-type *R*-branch transitions in the $v_t$ = 1 first excited torsional state of ${}^{13}CH_3OD$

| *J'* - *J''* | *K* | *TS* | $\nu_{obs}$ (MHz) | (o-c)[a] | $E_u$ ($cm^{-1}$)[a] | $\mu^2 S$[c] | *J'* - *J''* | *K* | *TS* | $\nu_{obs}$ (MHz) | (o-c)[a] | $E_u$ ($cm^{-1}$)[b] | $\mu^2 S$[c] |
|---|---|---|---|---|---|---|---|---|---|---|---|---|---|
| 1 - 0 | 0 | *A*+ | 44151.09[d] | 0.081 | 215.9972 | 0.71 | 7 - 6 | 0 | *A*+ | 309156.170 | -0.196 | 255.7681 | 4.91 |
| 1 - 0 | 0 | *E* | 44144.06[d] | 0.055 | 181.2842 | 0.71 | 7 - 6 | 1 | *A*+ | 306837.993 | 0.018 | 234.5803 | 4.94 |
| 4 - 3 | 0 | *A*+ | 176621.570 | -0.162 | 229.2526 | 2.80 | 7 - 6 | 1 | *A*– | 310682.025 | 0.031 | 235.0947 | 4.93 |
| 4 - 3 | 1 | *A*+ | 175464.453 | 0.042 | 208.2553 | 2.71 | 7 - 6 | 2 | *A*+ | 308525.522 | 0.030 | 225.5268 | 4.29 |
| 4 - 3 | 1 | *A*– | 177670.950 | 0.029 | 208.4394 | 2.70 | 7 - 6 | 2 | *A*– | 308664.368 | 0.108 | 225.5372 | 4.29 |
| 4 - 3 | 2 | *A*+ | 176397.498 | 0.020 | 199.0586 | 2.00 | 7 - 6 | 3 | *A*+ | 309122.846 | 0.017 | 250.5372 | 3.97 |
| 4 - 3 | 2 | *A*– | 176422.483 | 0.010 | 199.0598 | 2.00 | 7 - 6 | 3 | *A*– | 309134.487 | -0.018 | 250.5379 | 3.97 |
| 4 - 3 | 3 | *A*+ | 176579.710 | 0.044* | 224.0257 | 1.22 | 7 - 6 | 4 | *A*+ | 308753.624 | 0.019 | 299.2660 | 3.07 |
| 4 - 3 | 3 | *A*– | 176580.151 | -0.074 | 224.0257 | 1.22 | 7 - 6 | 4 | *A*– | 308753.624 | 0.030 | 299.2660 | 3.07 |
| 4 - 3 | 0 | *E* | 176531.362 | 0.031 | 194.5344 | 2.83 | 7 - 6 | 5 | *A*± | 307583.670 | 0.012 | 314.1749 | 1.61 |
| 4 - 3 | 1 | *E* | 176487.076 | 0.020 | 191.1411 | 2.59 | 7 - 6 | 6 | *A*± | 309829.777 | 0.089 | 322.2739 | 1.58 |
| 4 - 3 | 2 | *E* | 176561.834 | -0.071 | 221.2797 | 2.07 | 7 - 6 | 0 | *E* | 308758.165 | -0.003 | 221.0226 | 4.95 |
| 4 - 3 | 3 | *E* | 176897.324 | -0.021 | 254.8527 | 1.45 | 7 - 6 | 1 | *E* | 308717.063 | 0.008 | 217.6247 | 4.74 |
| 4 - 3 | –1 | *E* | 176571.001 | -0.087 | 225.5059 | 2.53 | 7 - 6 | 2 | *E* | 309012.880 | -0.060 | 247.7841 | 4.44 |
| 4 - 3 | –2 | *E* | 176661.196 | -0.016 | 230.3719 | 2.25 | 7 - 6 | 3 | *E* | 309015.554 | 0.012 | 281.3737 | 4.57 |
| 4 - 3 | –3 | *E* | 176244.288 | 0.004 | 218.3971 | 1.08 | 7 - 6 | 4 | *E* | 307731.569 | 0.024 | 274.9171 | 2.36 |
| 5 - 4 | 0 | *A*+ | 220790.289 | -0.174 | 236.6173 | 3.51 | 7 - 6 | 5 | *E* | 309679.923 | 0.058 | 287.0368 | 2.62 |
| 5 - 4 | 1 | *A*+ | 219286.286 | 0.029 | 215.5699 | 3.46 | 7 - 6 | 6 | *E* | 308908.156 | 0.083 | 333.2542 | 1.30 |
| 5 - 4 | 1 | *A*– | 222041.723 | 0.013 | 215.8459 | 3.46 | 7 - 6 | –1 | *E* | 309127.386 | -0.065 | 252.0172 | 4.61 |
| 5 - 4 | 2 | *A*+ | 220463.696 | 0.024 | 206.4125 | 2.80 | 7 - 6 | –2 | *E* | 308686.022 | -0.027 | 256.8624 | 4.77 |
| 5 - 4 | 2 | *A*– | 220513.547 | 0.001 | 206.4154 | 2.80 | 7 - 6 | –3 | *E* | 308248.450 | 0.003* | 244.8418 | 3.55 |
| 5 - 4 | 3 | *A*+ | 220746.184 | -0.040 | 231.3890 | 2.22 | 7 - 6 | –4 | *E* | 309501.250 | 0.053 | 263.2364 | 3.36 |
| 5 - 4 | 3 | *A*– | 220748.180 | 0.002 | 231.3891 | 2.22 | 7 - 6 | –5 | *E* | 308885.939 | 0.095 | 313.6347 | 2.35 |
| 5 - 4 | 4 | *A*+ | 220563.873 | -0.006 | 280.1389 | 1.17 | 7 - 6 | –6 | *E* | 307617.885 | 0.011 | 358.2657 | 0.86 |
| 5 - 4 | 4 | *A*– | 220563.873 | -0.005 | 280.1389 | 1.17 | 8 - 7 | 0 | *A*+ | 353356.892 | -0.201 | 267.5548 | 5.61 |
| 5 - 4 | 0 | *E* | 220630.717 | 0.036 | 201.8938 | 3.54 | 8 - 7 | 1 | *A*+ | 350557.828 | -0.014 | 246.2737 | 5.67 |
| 5 - 4 | 1 | *E* | 220582.463 | 0.006 | 198.4989 | 3.32 | 8 - 7 | 1 | *A*– | 354938.292 | 0.022 | 246.9341 | 5.65 |
| 5 - 4 | 2 | *E* | 220708.083 | -0.102 | 228.6418 | 2.90 | 8 - 7 | 2 | *A*+ | 352512.457 | 0.011 | 237.2853 | 5.01 |
| 5 - 4 | 3 | *E* | 221011.173 | -0.013 | 262.2248 | 2.62 | 8 - 7 | 2 | *A*– | 352719.717 | -0.044 | 237.3027 | 5.01 |
| 5 - 4 | 4 | *E* | 219790.760 | 0.011 | 255.8543 | 0.89 | 8 - 7 | 3 | *A*+ | 353336.722 | 0.025 | 262.3233 | 4.78 |
| 5 - 4 | –1 | *E* | 220739.002 | -0.089 | 232.8690 | 3.24 | 8 - 7 | 3 | *A*– | 353360.011 | 0.047 | 262.3247 | 4.78 |
| 5 - 4 | –2 | *E* | 220733.924 | -0.011 | 237.7348 | 3.14 | 8 - 7 | 4 | *A*+ | 352835.487 | 0.001 | 311.0353 | 3.91 |
| 5 - 4 | –3 | *E* | 220270.314 | -0.021* | 225.7446 | 1.98 | 8 - 7 | 4 | *A*– | 352835.487 | 0.031 | 311.0353 | 3.91 |
| 5 - 4 | –4 | *E* | 220881.784 | 0.005 | 244.0676 | 1.28 | 8 - 7 | 5 | *A*± | 351498.518 | 0.021 | 325.8996 | 2.31 |
| 6 - 5 | 0 | *A*+ | 264967.871 | -0.208 | 245.4557 | 4.21 | 8 - 7 | 6 | *A*± | 354037.358 | 0.083 | 334.0834 | 3.02 |
| 6 - 5 | 1 | *A*+ | 263079.246 | 0.074 | 224.3453 | 4.20 | 8 - 7 | 7 | *A*± | 353089.903 | -0.040 | 373.9206 | 0.04 |
| 6 - 5 | 1 | *A*– | 266380.935 | 0.013 | 224.7314 | 4.20 | 8 - 7 | 0 | *E* | 352777.530 | -0.026 | 232.7900 | 5.66 |
| 6 - 5 | 2 | *A*+ | 264507.821 | 0.025 | 215.2355 | 3.56 | 8 - 7 | 1 | *E* | 352749.236 | -0.025 | 229.3912 | 5.45 |
| 6 - 5 | 2 | *A*– | 264594.824 | -0.001 | 215.2413 | 3.56 | 8 - 7 | 2 | *E* | 353172.721 | -0.058 | 259.5646 | 5.18 |
| 6 - 5 | 3 | *A*+ | 264926.583 | 0.006 | 240.2260 | 3.13 | 8 - 7 | 3 | *E* | 352886.734 | 0.005 | 293.1447 | 5.42 |
| 6 - 5 | 3 | *A*– | 264931.772 | -0.007 | 240.2263 | 3.13 | 8 - 7 | 4 | *E* | 351706.035 | 0.038 | 286.6488 | 3.02 |
| 6 - 5 | 4 | *A*+ | 264662.618 | 0.003 | 288.9671 | 2.17 | 8 - 7 | 5 | *E* | 354090.854 | 0.080 | 298.8480 | 3.74 |
| 6 - 5 | 4 | *A*– | 264662.618 | 0.006 | 288.9671 | 2.17 | 8 - 7 | 6 | *E* | 353035.771 | 0.095 | 345.0302 | 2.44 |
| 6 - 5 | 5 | *A*± | 263659.755 | 0.004 | 303.9150 | 0.85 | 8 - 7 | 7 | *E* | 352589.788 | -0.243 | 410.0357 | 1.17 |
| 6 - 5 | 0 | *E* | 264707.782 | 0.023 | 210.7235 | 4.25 | 8 - 7 | –1 | *E* | 353353.202 | -0.053 | 263.8038 | 5.29 |
| 6 - 5 | 1 | *E* | 264660.328 | 0.038 | 207.3270 | 4.04 | 8 - 7 | –2 | *E* | 352545.550 | -0.017 | 268.6221 | 5.54 |
| 6 - 5 | 2 | *E* | 264858.213 | -0.082 | 237.4765 | 3.68 | 8 - 7 | –3 | *E* | 352191.904 | -0.001 | 256.5897 | 4.28 |
| 6 - 5 | 3 | *E* | 265054.201 | 0.001 | 271.0660 | 3.64 | 8 - 7 | –4 | *E* | 353901.461 | 0.068 | 275.0413 | 4.28 |
| 6 - 5 | 4 | *E* | 263759.457 | 0.015 | 264.6523 | 1.65 | 8 - 7 | –5 | *E* | 353003.373 | 0.079 | 325.4096 | 3.35 |
| 6 - 5 | 5 | *E* | 265325.290 | 0.060 | 276.7070 | 1.40 | 8 - 7 | –6 | *E* | 351549.812 | 0.025 | 369.9921 | 1.63 |
| 6 - 5 | –1 | *E* | 264923.490 | -0.093 | 241.7059 | 3.93 | 8 - 7 | –7 | *E* | 353463.8 | -0.078* | 380.3073 | 1.49 |
| 6 - 5 | –2 | *E* | 264745.757 | -0.026 | 246.5658 | 3.98 | 9 - 8 | 0 | *A*+ | 397571.780 | -0.215 | 280.8164 | 6.31 |
| 6 - 5 | –3 | *E* | 264273.313 | 0.121 | 234.5598 | 2.79 | 9 - 8 | 1 | *A*+ | 394234.356 | -0.036 | 259.4239 | 6.39 |
| 6 - 5 | –4 | *E* | 265163.631 | 0.040 | 252.9125 | 2.37 | 9 - 8 | 1 | *A*– | 399143.064 | 0.031 | 260.2481 | 6.37 |
| 6 - 5 | –5 | *E* | 264765.255 | 0.055 | 303.3314 | 1.26 | 9 - 8 | 2 | *A*+ | 396464.444 | 0.038 | 250.5100 | 5.72 |

Table 2 (*continued*)

| J′ - J″ | K | TS | $\nu_{obs}$ (MHz) | (o-c)[a] | $E_u$ (cm$^{-1}$)[a] | $\mu^2S$[c] | J′ - J″ | K | TS | $\nu_{obs}$ (MHz) | (o-c)[a] | $E_u$ (cm$^{-1}$)[b] | $\mu^2S$[c] |
|---|---|---|---|---|---|---|---|---|---|---|---|---|---|
| 9 - 8 | 2 | *A*– | 396759.128 | -0.069 | 250.5372 | 5.72 | 10 - 9 | 8 | *E* | 439802.598 | -0.114 | 464.1806 | 2.05 |
| 9 - 8 | 3 | *A*+ | 397569.506 | 0.072* | 275.5848 | 5.56 | 10 - 9 | 9 | *E* | 441376.959 | 0.057 | 492.7746 | 1.44 |
| 9 - 8 | 3 | *A*– | 397611.938 | 0.062 | 275.5876 | 5.56 | 10 - 9 | –1 | *E* | 441878.481 | 0.015 | 291.8059 | 6.62 |
| 9 - 8 | 4 | *A*+ | 396906.874 | 0.010 | 324.2747 | 4.72 | 10 - 9 | –2 | *E* | 439997.886 | -0.055 | 296.5186 | 7.02 |
| 9 - 8 | 4 | *A*– | 396906.874 | 0.082 | 324.2747 | 4.72 | 10 - 9 | –3 | *E* | 439968.273 | -0.070 | 284.4779 | 5.70 |
| 9 - 8 | 5 | *A*± | 395402.917 | 0.030 | 339.0889 | 2.97 | 10 - 9 | –4 | *E* | 442905.312 | 0.073 | 303.1031 | 6.02 |
| 9 - 8 | 6 | *A*± | 398223.947 | 0.096 | 347.3667 | 4.29 | 10 - 9 | –5 | *E* | 441226.788 | 0.070 | 353.3738 | 5.15 |
| 9 - 8 | 7 | *A*± | 397229.377 | -0.043 | 387.1708 | 0.59 | 10 - 9 | –6 | *E* | 439392.851 | 0.023 | 397.8403 | 3.03 |
| 9 - 8 | 8 | *A*± | 396932.789 | 0.156* | 457.2232 | | 10 - 9 | –7 | *E* | 441603.887 | -0.012 | 408.2984 | 4.02 |
| 9 - 8 | 0 | *E* | 396761.643 | 0.005 | 246.0245 | 6.36 | 10 - 9 | –8 | *E* | 441440.872 | -0.266 | 441.6290 | 2.65 |
| 9 - 8 | 1 | *E* | 396753.411 | -0.027 | 242.6254 | 6.15 | 10 - 9 | –9 | *E* | 441063.849 | 0.310 | 508.8864 | 1.33 |
| 9 - 8 | 2 | *E* | 397338.355 | -0.034 | 272.8184 | 5.91 | 11 - 10 | 0 | *A*+ | 486050.787 | -0.162 | 311.7662 | 7.72 |
| 9 - 8 | 3 | *E* | 396662.402 | 0.048 | 306.3760 | 6.21 | 11 - 10 | 1 | *A*+ | 481442.548 | -0.148 | 290.0887 | 7.82 |
| 9 - 8 | 4 | *E* | 395681.289 | 0.032 | 299.8473 | 3.67 | 11 - 10 | 1 | *A*– | 487371.226 | 0.040 | 291.2916 | 7.78 |
| 9 - 8 | 5 | *E* | 398560.515 | 0.078 | 312.1425 | 4.79 | 11 - 10 | 2 | *A*+ | 484246.791 | 0.016 | 281.3521 | 7.12 |
| 9 - 8 | 6 | *E* | 397162.471 | 0.088 | 358.2781 | 3.49 | 11 - 10 | 2 | *A*– | 484781.004 | -0.143 | 281.4106 | 7.12 |
| 9 - 8 | 7 | *E* | 396636.766 | -0.244 | 423.2661 | 2.22 | 11 - 10 | 3 | *A*+ | 486093.813 | 0.071 | 306.5367 | 7.08 |
| 9 - 8 | 8 | *E* | 395876.410 | -0.090* | 449.5103 | 1.07 | 11 - 10 | 3 | *A*– | 486210.291 | 0.167 | 306.5458 | 7.08 |
| 9 - 8 | –1 | *E* | 397603.080 | -0.018 | 277.0665 | 5.95 | 11 - 10 | 4 | *A*+ | 485012.291 | -0.067* | 355.1621 | 6.25 |
| 9 - 8 | –2 | *E* | 396317.415 | -0.045 | 281.8418 | 6.29 | 11 - 10 | 4 | *A*– | 485012.291 | 0.244* | 355.1620 | 6.25 |
| 9 - 8 | –3 | *E* | 396099.630 | -0.055 | 269.8022 | 4.99 | 11 - 10 | 5 | *A*± | 483174.714 | 0.085 | 369.8591 | 4.23 |
| 9 - 8 | –4 | *E* | 398368.899 | 0.073 | 288.3294 | 5.16 | 11 - 10 | 6 | *A*± | 486525.049 | 0.097 | 378.3518 | 6.57 |
| 9 - 8 | –5 | *E* | 397117.166 | 0.088 | 338.6560 | 4.27 | 11 - 10 | 7 | *A*± | 485510.782 | -0.125 | 418.0882 | 4.66 |
| 9 - 8 | –6 | *E* | 395475.266 | 0.028 | 383.1838 | 2.35 | 11 - 10 | 8 | *A*± | 485089.944 | 0.020 | 488.1147 | 2.88 |
| 9 - 8 | –7 | *E* | 397550.046 | -0.032 | 393.5681 | 2.81 | 11 - 10 | 9 | *A*± | 483128.185 | -0.005 | 542.9836 | |
| 9 - 8 | –8 | *E* | 397308.488 | -0.217 | 426.9042 | 1.39 | 11 - 10 | 10 | *A*± | 484919.021 | 0.884* | 571.3626 | |
| 10 - 9 | 0 | *A*+ | 441802.571 | -0.180 | 295.5533 | 7.02 | 11 - 10 | 0 | *E* | 484607.064 | -0.132 | 276.8896 | 7.77 |
| 10 - 9 | 1 | *A*+ | 437863.654 | -0.110 | 274.0295 | 7.11 | 11 - 10 | 1 | *E* | 484663.865 | -0.096 | 273.4931 | 7.55 |
| 10 - 9 | 1 | *A*– | 443289.582 | 0.027 | 275.0347 | 7.08 | 11 - 10 | 2 | *E* | 485688.662 | 0.017 | 303.7464 | 7.35 |
| 10 - 9 | 2 | *A*+ | 440377.212 | 0.011 | 265.1994 | 6.42 | 11 - 10 | 3 | *E* | 483924.377 | 0.113 | 337.2061 | 7.66 |
| 10 - 9 | 2 | *A*– | 440780.291 | -0.104 | 265.2400 | 6.42 | 11 - 10 | 4 | *E* | 483626.146 | 0.118 | 330.6446 | 4.95 |
| 10 - 9 | 3 | *A*+ | 441821.791 | 0.041 | 290.3223 | 6.33 | 11 - 10 | 5 | *E* | 487669.210 | 0.115 | 343.1893 | 6.76 |
| 10 - 9 | 3 | *A*– | 441894.151 | 0.136 | 290.3276 | 6.33 | 11 - 10 | 6 | *E* | 485412.639 | 0.148 | 389.1895 | 5.40 |
| 10 - 9 | 4 | *A*+ | 440966.300 | -0.010 | 338.9838 | 5.49 | 11 - 10 | 7 | *E* | 484700.714 | -0.319 | 454.1333 | 4.09 |
| 10 - 9 | 4 | *A*– | 440966.300 | 0.146 | 338.9838 | 5.49 | 11 - 10 | 8 | *E* | 483709.974 | -0.110* | 480.3154 | 2.95 |
| 10 - 9 | 5 | *A*± | 439295.471 | 0.056 | 353.7422 | 3.60 | 11 - 10 | 9 | *E* | 485388.350 | 0.226 | 508.9654 | 2.75 |
| 10 - 9 | 6 | *A*± | 442387.220 | 0.087 | 362.1231 | 5.47 | 11 - 10 | 10 | *E* | 485187.680 | -0.359 | 570.4450 | 1.37 |
| 10 - 9 | 7 | *A*± | 441369.682 | -0.082 | 401.8933 | 3.12 | 11 - 10 | –1 | *E* | 486180.149 | 0.113 | 308.0232 | 7.27 |
| 10 - 9 | 8 | *A*± | 441014.708 | 0.121 | 471.9339 | 2.41 | 11 - 10 | –2 | *E* | 483586.962 | -0.055 | 312.6493 | 7.74 |
| 10 - 9 | 9 | *A*± | 439243.901 | -0.005 | 526.8682 | 0.94 | 11 - 10 | –3 | *E* | 483794.897 | -0.095 | 300.6156 | 6.39 |
| 10 - 9 | 0 | *E* | 440706.164 | -0.050 | 260.7249 | 7.07 | 11 - 10 | –4 | *E* | 487508.947 | 0.111 | 319.3647 | 6.85 |
| 10 - 9 | 1 | *E* | 440726.067 | -0.074 | 257.3265 | 6.85 | 11 - 10 | –5 | *E* | 485331.819 | 0.095 | 369.5627 | 6.00 |
| 10 - 9 | 2 | *E* | 441510.225 | -0.015* | 287.5456 | 6.63 | 11 - 10 | –6 | *E* | 483301.088 | 0.086 | 413.9615 | 3.68 |
| 10 - 9 | 3 | *E* | 440340.800 | 0.072 | 321.0641 | 6.95 | 11 - 10 | –7 | *E* | 485622.686 | 0.047 | 424.4971 | 5.13 |
| 10 - 9 | 4 | *E* | 439655.450 | 0.083 | 314.5126 | 4.32 | 11 - 10 | –8 | *E* | 485568.807 | -0.289 | 457.8259 | 3.81 |
| 10 - 9 | 5 | *E* | 443088.178 | 0.124 | 326.9224 | 5.79 | 11 - 10 | –9 | *E* | 485150.636 | 0.190 | 525.0692 | 2.54 |
| 10 - 9 | 6 | *E* | 441288.135 | 0.092 | 372.9979 | 4.47 | 11 - 10 | –10 | *E* | 484058.442 | -0.130 | 604.7657 | 1.11 |
| 10 - 9 | 7 | *E* | 440674.105 | -0.286* | 437.9654 | 3.18 | | | | | | | |

[a] Observed minus calculated line frequencies in MHz. Asterisks indicate blended or barely resolved lines that were excluded from the fit.

[b] Calculated upper level energies in cm$^{-1}$. The zero reference energy is taken as the $J = K = 0$ $A$+ $v_t = 0$ level, calculated to lie 104.8883 cm$^{-1}$ above the bottom of the torsional barrier.

[c] $S$ is the transition strength. Relative line intensities are given by the product of $\mu^2S$ and the appropriate Boltzmann factor. The dipole moment components were taken as $\mu_a = 0.836$ and $\mu_b = -1.439$ Debye.

## Table 3

Millimeter-wave frequencies, upper level energies ($E_u$) and line strengths ($\mu^2S$) for $^rQ_K(J)$ and $^pQ_K(J)$ *b*-type *Q*-branch transitions in the $v_t = 0$ ground torsional state of $^{13}CH_3OD$

| $J$ | $\nu_{obs}$ (MHz) | (o-c)[a] | $E_u$ (cm$^{-1}$)[b] | $\mu^2S$[c] |
|---|---|---|---|---|
| $K = 1A^- \leftarrow 0A^+$ | | | | |
| 7 | 152780.795 | 0.067 | 5.9577 | 13.97 |
| 8 | 158567.871 | 0.055 | 8.9523 | 15.45 |
| 9 | 165251.534 | 0.067 | 13.4437 | 16.80 |
| 10 | 172885.520 | 0.048 | 19.4317 | 17.99 |
| 11 | 181524.336 | 0.043 | 26.9155 | 19.03 |
| 12 | 191220.711 | 0.026 | 35.8945 | 19.91 |
| 13 | 202022.975 | 0.029 | 46.3676 | 20.62 |
| 14 | 213971.897 | 0.002 | 58.3339 | 21.16 |
| 15 | 227097.754 | -0.008 | 71.7921 | 21.56 |
| 16 | 241417.151 | -0.047 | 86.7406 | 21.80 |
| 17 | 256930.602 | -0.055 | 103.1778 | 21.92 |
| 18 | 273620.311 | -0.068 | 121.1018 | 21.92 |
| 19 | 291449.091 | -0.080 | 287.9835 | 21.83 |
| 20 | 310360.016 | -0.100 | 317.7004 | 21.67 |
| 21 | 330277.197 | -0.078 | 348.8742 | 21.45 |
| 22 | 351107.268 | -0.089 | 381.5002 | 21.19 |
| 23 | 372742.215 | -0.085 | 415.5733 | 20.91 |
| 24 | 395062.585 | -0.049 | 451.0885 | 20.61 |
| 25 | 417941.397 | -0.060 | 488.0405 | 20.31 |
| $K = 2A^+ \leftarrow 1A^-$ | | | | |
| 2 | 320451.125 | -0.055 | 19.6414 | 1.68 |
| 3 | 318546.911 | -0.062 | 24.0693 | 2.97 |
| 4 | 316061.217 | -0.067 | 29.9744 | 4.19 |
| 5 | 313039.831 | -0.065 | 37.3574 | 5.41 |
| 6 | 309540.044 | -0.060 | 46.2196 | 6.66 |
| 7 | 305630.534 | -0.072 | 56.5624 | 7.97 |
| 8 | 301391.117 | -0.101 | 68.3873 | 9.35 |
| 9 | 296912.264 | -0.089 | 81.6960 | 10.81 |
| 10 | 292294.115 | -0.085 | 96.4905 | 12.35 |
| 11 | 287645.468 | -0.087 | 112.7726 | 13.97 |
| 12 | 283082.199 | -0.096 | 130.5444 | 15.68 |
| 13 | 278725.426 | -0.099 | 149.8077 | 17.45 |
| 14 | 274699.384 | -0.091 | 170.5642 | 19.29 |
| 15 | 271129.178 | -0.102 | 192.8154 | 21.16 |
| 16 | 268138.704 | -0.079 | 216.5625 | 23.05 |
| 17 | 265848.452 | -0.092 | 241.8061 | 24.92 |
| 18 | 264374.043 | -0.108 | 268.5467 | 26.75 |
| 19 | 263824.812 | -0.093 | 296.7838 | 28.50 |
| 20 | 264302.775 | -0.082 | 326.5166 | 30.13 |
| 21 | 265901.987 | -0.079 | 357.7437 | 31.62 |
| 22 | 268707.859 | -0.057 | 390.4633 | 32.94 |
| 23 | 272796.287 | 0.002 | 424.6728 | 34.07 |
| 24 | 278232.466 | 0.063 | 460.3693 | 34.99 |
| 25 | 285069.484 | 0.206 | 497.5494 | 35.68 |
| $K = 2A^- \leftarrow 1A^+$ | | | | |
| 2 | 324324.331 | -0.053 | 19.6410 | 1.65 |
| 3 | 326260.855 | -0.045 | 24.0675 | 2.87 |
| 4 | 328845.393 | -0.029 | 29.9690 | 3.96 |
| 5 | 332079.814 | -0.022 | 37.3449 | 4.98 |
| 6 | 335966.196 | 0.006 | 46.1947 | 5.94 |
| 7 | 340506.510 | 0.020 | 56.5177 | 6.86 |
| 8 | 345702.479 | 0.007 | 68.3129 | 7.72 |
| 9 | 351555.393 | 0.040 | 81.5795 | 8.55 |
| 10 | 358065.585 | 0.034 | 96.3163 | 9.33 |
| 11 | 365232.481 | 0.065 | 112.5222 | 10.06 |
| $K = 2A^- \leftarrow 1A^+$, cont'd | | | | |
| 12 | 373053.968 | 0.028 | 130.1959 | 10.74 |
| 13 | 381526.580 | 0.088 | 149.3359 | 11.37 |
| 14 | 390644.674 | 0.114 | 169.9407 | 11.96 |
| 15 | 400400.626 | 0.093 | 192.0087 | 12.49 |
| 16 | 410784.647 | 0.137 | 215.5381 | 12.98 |
| 17 | 421784.263 | 0.097 | 240.5272 | 13.43 |
| 18 | 433384.580 | -0.085 | 266.9739 | 13.82 |
| 19 | 445568.751 | 0.117 | 294.8764 | 14.18 |
| 20 | 458316.288 | 0.087 | 324.2325 | 14.49 |
| 21 | 471605.150 | 0.058 | 355.0401 | 14.77 |
| 22 | 485410.822 | 0.031 | 387.2969 | 15.01 |
| 23 | 499706.763 | 0.012 | 421.0009 | 15.21 |
| $K = 3A^- \leftarrow 2A^+$ | | | | |
| 3 | 403358.623 | 0.070 | 37.5239 | 1.75 |
| 4 | 403279.799 | 0.053 | 43.4263 | 3.15 |
| 5 | 403130.189 | 0.035 | 50.8044 | 4.40 |
| 6 | 402876.569 | 0.034 | 59.6581 | 5.59 |
| 7 | 402480.043 | 0.009 | 69.9877 | 6.74 |
| 8 | 401897.006 | 0.007 | 81.7931 | 7.86 |
| 9 | 401080.134 | -0.018 | 95.0746 | 8.99 |
| 10 | 399980.113 | -0.031 | 109.8324 | 10.11 |
| 11 | 398547.526 | -0.030 | 126.0667 | 11.25 |
| 12 | 396735.305 | -0.031 | 143.7781 | 12.40 |
| 13 | 394501.612 | -0.027 | 162.9668 | 13.58 |
| 14 | 391812.903 | -0.049 | 183.6336 | 14.80 |
| 15 | 388647.310 | -0.026 | 205.7793 | 16.06 |
| 16 | 384997.569 | -0.010 | 229.4046 | 17.36 |
| 17 | 380873.992 | -0.018 | 254.5107 | 18.73 |
| 18 | 376306.748 | 0.010 | 281.0989 | 20.17 |
| 19 | 371347.154 | 0.013 | 309.1706 | 21.69 |
| 20 | 366068.448 | 0.006 | 338.7273 | 23.30 |
| 21 | 360565.355 | 0.011 | 369.7709 | 25.00 |
| 22 | 354952.706 | 0.007 | 402.3033 | 26.80 |
| 23 | 349363.257 | -0.035 | 436.3263 | 28.70 |
| 24 | 343944.798 | -0.046 | 471.8421 | 30.69 |
| 25 | 338856.318 | -0.064 | 508.8524 | 32.75 |
| $K = 3A^+ \leftarrow 2A^-$ | | | | |
| 3 | 403411.877 | 0.072 | 37.5239 | 1.75 |
| 4 | 403439.164 | 0.062 | 43.4263 | 3.15 |
| 5 | 403500.813 | 0.045 | 50.8043 | 4.40 |
| 6 | 403614.697 | 0.018 | 59.6579 | 5.58 |
| 7 | 403801.814 | 0.007 | 69.9871 | 6.71 |
| 8 | 404085.965 | 0.003 | 81.7918 | 7.82 |
| 9 | 404493.487 | -0.012 | 95.0719 | 8.91 |
| 10 | 405052.950 | -0.040 | 109.8274 | 9.99 |
| 11 | 405794.833 | -0.036 | 126.0581 | 11.05 |
| 12 | 406750.972 | -0.073 | 143.7636 | 12.11 |
| 13 | 407954.420 | -0.084 | 162.9438 | 13.14 |
| 14 | 409438.815 | -0.075 | 183.5981 | 14.17 |
| 15 | 411237.979 | -0.101 | 205.7261 | 15.17 |
| 16 | 413385.655 | -0.101 | 229.3272 | 16.15 |
| 17 | 415914.899 | -0.080 | 254.4006 | 17.11 |
| 18 | 418857.694 | -0.082 | 280.9455 | 18.04 |
| 19 | 422244.658 | -0.079 | 308.9610 | 18.94 |
| 20 | 426104.576 | -0.058 | 338.4458 | 19.80 |
| $K = 3A^+ \leftarrow 2A^-$, cont'd | | | | |
| 21 | 430464.002 | -0.057 | 369.3988 | 20.62 |
| 22 | 435347.073 | -0.018 | 401.8186 | 21.39 |
| 23 | 440774.998 | 0.000 | 435.7035 | 22.12 |
| 24 | 446765.991 | 0.031 | 471.0521 | 22.80 |
| 25 | 453334.906 | 0.060 | 507.8623 | 23.43 |
| $K = 2 \leftarrow 1\ E$ | | | | |
| 2 | 198428.017 | 0.041 | 17.3410 | 1.68 |
| 3 | 198360.778 | 0.034 | 21.7688 | 2.96 |
| 4 | 198195.135 | 0.026 | 27.6732 | 4.14 |
| 5 | 197870.877 | 0.010 | 35.0547 | 5.29 |
| 6 | 197323.503 | -0.012 | 43.9141 | 6.45 |
| 7 | 196497.405 | -0.001 | 54.2528 | 7.67 |
| 8 | 195364.774 | -0.007 | 66.0725 | 8.99 |
| 9 | 193948.246 | -0.013 | 79.3763 | 10.48 |
| 10 | 192339.981 | -0.017 | 94.1681 | 12.20 |
| 11 | 190705.599 | -0.015 | 110.4536 | 14.18 |
| 12 | 189261.203 | -0.009 | 128.2385 | 16.36 |
| 13 | 188228.598 | -0.016 | 147.5279 | 18.68 |
| 14 | 187797.048 | -0.025 | 168.3252 | 21.03 |
| 15 | 188113.481 | -0.044 | 190.6315 | 23.37 |
| 16 | 189293.385 | -0.062 | 214.4465 | 25.64 |
| 17 | 191433.934 | -0.080 | 239.7690 | 27.82 |
| 18 | 194621.819 | -0.093 | 266.5974 | 29.87 |
| 19 | 198936.654 | -0.106 | 294.9294 | 31.75 |
| 20 | 204452.295 | -0.108 | 324.7624 | 33.44 |
| 21 | 211237.176 | -0.098 | 356.0937 | 34.90 |
| 22 | 219354.018 | -0.065 | 388.9199 | 36.12 |
| 23 | 228858.851 | -0.005 | 423.2378 | 37.10 |
| 24 | 239799.384 | 0.102 | 459.0437 | 37.83 |
| 25 | 252212.751 | 0.258 | 496.3339 | 38.32 |
| $K = 3 \leftarrow 2\ E$ | | | | |
| 3 | 405822.066 | -0.081 | 35.3056 | 1.79 |
| 4 | 405789.983 | -0.055 | 41.2089 | 3.24 |
| 5 | 405728.461 | -0.073 | 48.5883 | 4.54 |
| 6 | 405619.068 | -0.069 | 57.4441 | 5.77 |
| 7 | 405432.287 | -0.086 | 67.7765 | 6.96 |
| 8 | 405120.302 | -0.079 | 79.5859 | 8.11 |
| 9 | 404606.913 | -0.079 | 92.8725 | 9.19 |
| 10 | 403778.714 | -0.084 | 107.6367 | 10.14 |
| 11 | 402486.268 | -0.070 | 123.8791 | 10.89 |
| 12 | 400566.229 | -0.087 | 141.5999 | 11.39 |
| 13 | 397881.180 | -0.077 | 160.7998 | 11.66 |
| 14 | 394350.780 | -0.089 | 181.4793 | 11.81 |
| 15 | 389955.015 | -0.059 | 203.6390 | 11.93 |
| 16 | 384717.624 | -0.079 | 227.2793 | 12.07 |
| 17 | 378688.828 | -0.025 | 252.4007 | 12.27 |
| 18 | 371933.772 | 0.001 | 279.0038 | 12.51 |
| 19 | 364525.375 | 0.037 | 307.0887 | 12.77 |
| 20 | 356538.474 | 0.056 | 336.6553 | 13.01 |
| 21 | 348043.686 | 0.044 | 367.7031 | 13.17 |
| 22 | 339100.440 | 0.082 | 400.2311 | 13.20 |
| 23 | 329749.119 | 0.057 | 434.2370 | 13.02 |
| 24 | 320004.548 | 0.023 | 469.7179 | 12.57 |
| 25 | 309851.912 | -0.054 | 506.6694 | 11.82 |

Table 3 (*continued*)

| $J$ | $\nu_{obs}$ (MHz) | (o-c)[a] | $E_u$ (cm$^{-1}$)[b] | $\mu^2S$[c] | $J$ | $\nu_{obs}$ (MHz) | (o-c)[a] | $E_u$ (cm$^{-1}$)[b] | $\mu^2S$[c] | $J$ | $\nu_{obs}$ (MHz) | (o-c)[c] | $E_u$ (cm$^{-1}$)[b] | $\mu^2S$[c] |
|---|---|---|---|---|---|---|---|---|---|---|---|---|---|---|
| $K = -2 \leftarrow -1$ $E$ | | | | | $K = -2 \leftarrow -1$ $E$, cont'd | | | | | $K = -2 \leftarrow 0$ $E$, cont'd | | | | |
| 2 | 270973.310 | -0.020 | 16.7212 | 1.71 | 18 | 350952.86 | -0.68 | 237.9323 | D | 20 | 378715.734 | 0.077 | 321.6851 | 9.05 |
| 3 | 271096.615 | -0.005 | 21.1493 | 3.01 | 19 | 362048.19 | -0.81 | 264.3957 | D | 21 | 391406.723 | 0.001 | 352.5080 | 10.28 |
| 4 | 271425.234 | 0.001 | 27.0542 | 4.19 | 20 | 373812.25 | -0.91 | 292.3138 | D | 22 | 404822.804 | 0.056 | 384.7805 | 11.17 |
| 5 | 272094.393 | 0.047 | 34.4366 | 5.33 | 21 | 386282.59 | -1.01 | 321.6851 | D | $K = -3 \leftarrow -2$ $E$ | | | | |
| 6 | 273261.097 | 0.076 | 43.2972 | 6.44 | 22 | 399471.17 | -1.02 | 352.5080 | D | 3 | 510101.191 | 0.077 | 38.1645 | 1.76 |
| 7 | 275089.882 | 0.101 | 53.6363 | 7.50 | 23 | 413364.70 | -1.03 | 264.3957 | D | 4 | 510032.518 | 0.001 | 44.0671 | 3.18 |
| 8 | 277731.712 | 0.120 | 65.4541 | 8.52 | $K = -2 \leftarrow 0$ $E$ | | | | | 5 | 509911.900 | 0.056 | 51.4454 | 4.46 |
| 9 | 281298.656 | 0.131 | 78.7496 | 9.47 | 11 | 305267.60 | -0.81 | 109.7636 | D | 6 | 509721.208 | 0.039 | 60.2996 | 5.65 |
| 10 | 285840.905 | 0.105 | 93.5207 | 10.34 | 12 | 309988.12 | -0.94 | 127.4739 | D | 7 | 509448.099 | 0.018 | 70.6297 | 6.80 |
| 11 | 291338.278 | 0.086 | 109.7636 | 11.09 | 13 | 315456.86 | -0.86 | 146.6476 | D | 8 | 509094.938 | 0.010 | 82.4357 | 7.90 |
| 12 | 297717.740 | 0.061 | 127.4739 | 11.65 | 14 | 321708.44 | -0.85 | 167.2816 | D | 9 | 508691.427 | 0.003 | 95.7177 | 8.92 |
| 13 | 304892.609 | 0.040 | 146.6476 | 12.00 | 15 | 328806.60 | -0.79 | 189.3743 | D | 10 | 508307.138 | -0.038 | 110.4760 | 9.79 |
| 14 | 312795.241 | -0.045 | 167.2816 | 12.11 | 16 | 336825.71 | -0.64 | 212.9248 | D | 11 | 508054.824 | -0.032 | 126.7105 | 10.45 |
| 15 | 321382.229 | -0.155 | 189.3743 | 11.91 | 17 | 345828.45 | | 237.9323 | D | 12 | 508072.756 | -0.068 | 144.4214 | 10.84 |
| 16 | 330621.245 | -0.284 | 212.9248 | 11.30 | 18 | 355840.471 | -0.226 | 264.3957 | 5.55 | 13 | 508490.170 | -0.102 | 163.6090 | 10.99 |
| 17 | 340483.13 | -0.46 | 237.9323 | d | 19 | 366829.627 | -0.112 | 292.3138 | 7.43 | 14 | 509399.902 | -0.165 | 184.2733 | 10.98 |

[a] Observed minus calculated line frequencies in MHz.

[b] Calculated upper level energies in cm$^{-1}$. The zero reference energy is taken as the $J = K = 0$ $A$+ $v_t = 0$ level, calculated to lie 104.8883 cm$^{-1}$ above the bottom of the torsional barrier.

[c] $S$ is the transition strength. Relative line intensities are given by the product of $\mu^2S$ and the appropriate Boltzmann factor. The dipole moment components were taken as $\mu_a = 0.836$ and $\mu_b = -1.439$ Debye.

[d] For $J \geq 17$, the $K$-labeling for the strongly mixed $K = 0$ and $-1$ $E$ $v_t = 0$ levels was interchanged by the computer program and the $J = 17$ $K = 0$ $E$ level was missing. Thus, the line strengths were not calculated for $Q$-branch lines with $J \geq 17$ for $K = -2 \leftarrow -1$ and with $J \leq 17$ for $K = -2 \leftarrow 0$. These transitions were not included in the fit but were located later.

**Table 4**

Millimeter-wave frequencies, upper level energies ($E_u$) and line strengths ($\mu^2S$) for ${}^{r}Q_K(J)$ and ${}^{p}Q_K(J)$ *b*-type *Q*-branch transitions in the $v_t = 1$ first excited torsional state of ${}^{13}CH_3OD$

| *J* | $\nu_{obs}$ (MHz) | (o-c)[a] | $E_u$ ($cm^{-1}$)[b] | $\mu^2S$[c] | *J* | $\nu_{obs}$ (MHz) | (o-c)[a] | $E_u$ ($cm^{-1}$)[b] | $\mu^2S$[c] | *J* | $\nu_{obs}$ (MHz) | (o-c)[c] | $E_u$ ($cm^{-1}$)[b] | $\mu^2S$[c] |
|---|---|---|---|---|---|---|---|---|---|---|---|---|---|---|
| $K = 1A^+ \leftarrow 2A^-$ | | | | | $K = 5A^- \leftarrow 4A^+$, cont'd | | | | | $K = 6A^- \leftarrow 5A^+$, cont'd | | | | |
| 2 | 277336.221 | -0.022 | 198.0121 | 1.58 | 10 | 442445.402 | 0.043* | 353.7422 | 4.93 | 22 | 304987.539 | -0.165 | 653.7402 | 14.24 |
| 3 | 276631.757 | 0.026 | 202.4025 | 2.73 | 11 | 440607.841 | 0.212* | 369.8591 | 5.51 | 23 | 310545.176 | -0.177 | 687.5520 | 14.66 |
| 4 | 275673.717 | 0.048 | 208.2553 | 3.74 | 12 | 438603.041 | -0.123 | 387.4393 | 6.05 | 24 | 316249.855 | -0.162 | 722.8215 | 15.08 |
| 5 | 274446.456 | 0.076 | 215.5699 | 4.67 | 13 | 436431.983 | -0.092 | 406.4820 | 6.56 | 25 | 322107.973 | 0.243 | 759.5474 | 15.50 |
| 6 | 272930.823 | 0.096 | 224.3453 | 5.53 | 14 | 434094.402 | -0.079 | 426.9868 | 7.02 | | | | | |
| 7 | 271104.559 | 0.117 | 234.5803 | 6.32 | 15 | 431590.395 | -0.108 | 448.9529 | 7.45 | $K = 6A^+ \leftarrow 5A^-$ | | | | |
| 8 | 268942.645 | 0.122 | 246.2737 | 7.04 | 16 | 428920.192 | -0.069 | 472.3797 | 7.84 | 16[d] | 274784.862 | -0.137 | 481.5455 | 11.40 |
| 9 | 266417.872 | 0.154 | 259.4239 | 7.69 | 17 | 426083.786 | -0.072 | 497.2664 | 8.20 | 17 | 279417.144 | -0.241 | 506.5867 | 11.95 |
| 10 | 263501.239 | 0.152 | 274.0295 | 8.27 | 18 | 423081.284 | -0.089 | 523.6121 | 8.52 | 18 | 284220.973 | -0.386 | 533.0927 | 12.45 |
| 11 | 260162.743 | 0.108 | 290.0887 | 8.77 | 19 | 419912.748 | -0.098 | 551.4160 | 8.81 | 19 | 289187.396 | -0.090 | 561.0623 | 12.93 |
| 12 | 256372.077 | 0.058 | 307.5997 | 9.21 | 20 | 416578.107 | -0.156 | 580.6771 | 9.07 | 20 | 294307.943 | -0.131 | 590.4942 | 13.38 |
| 13 | 252099.285 | -0.003 | 326.5608 | 9.56 | 21 | 413077.340 | -0.202 | 611.3945 | 9.30 | 21 | 299577.435 | -0.123 | 621.3873 | 13.82 |
| 14 | 247315.596 | -0.071 | 346.9701 | 9.85 | 22 | 409410.250 | -0.267 | 643.5669 | 9.49 | 22 | 304992.855 | -0.046 | 653.7404 | 14.24 |
| 15 | 241994.185 | -0.134 | 368.8257 | 10.06 | 23 | 405576.438 | -0.480 | 677.1934 | 9.66 | 23 | 310554.020 | 0.016 | 687.5523 | 14.66 |
| 16 | 236110.893 | -0.199 | 392.1257 | 10.21 | 24 | 401575.892 | -0.469 | 712.2726 | 9.79 | 24 | 316264.236 | 0.112 | 722.8220 | 15.08 |
| 17 | 229644.945 | -0.241 | 416.8682 | 10.29 | 25 | 397407.729 | -0.601 | 748.8032 | 9.89 | 25 | 322130.475 | 0.182 | 759.5483 | 15.51 |
| 18 | 222579.480 | -0.250 | 443.0516 | 10.32 | $K = 5A^+ \leftarrow 4A^-$ | | | | | $K = 3 \leftarrow 4\ E$ | | | | |
| 19 | 214902.014 | -0.224 | 470.6741 | 10.29 | 5 | 449130.007 | -0.026 | 295.1203 | 1.10 | 4 | 189763.612 | 0.049 | 254.8527 | 1.39 |
| 20 | 206604.840 | -0.087 | 499.7341 | 10.22 | 6 | 448127.146 | -0.026 | 303.9150 | 2.05 | 5 | 190984.038 | 0.039 | 262.2248 | 2.41 |
| 21 | 197684.949 | 0.061 | 530.2300 | 10.11 | 7 | 446957.179 | -0.057 | 314.1749 | 2.88 | 6 | 192278.787 | 0.029 | 271.0660 | 3.17 |
| 22 | 188144.491 | 0.377 | 562.1604 | 9.97 | 8 | 445620.211 | -0.065 | 325.8996 | 3.62 | 7 | 193562.755 | 0.001 | 281.3737 | 3.73 |
| 23 | 177990.204 | 0.795 | 595.5240 | 9.80 | 9 | 444116.231 | -0.140 | 339.0889 | 4.30 | 8 | 194743.501 | 0.016 | 293.1447 | 4.11 |
| 24 | 167233.677 | 0.042 | 630.3196 | 9.59 | 10 | 442445.402 | -0.229* | 353.7422 | 4.93 | 9 | 195724.570 | -0.012 | 306.3760 | 4.33 |
| 25 | 155887.090 | -0.416 | 666.5458 | 9.40 | 11 | 440607.841 | -0.372* | 369.8591 | 5.51 | 10 | 196409.920 | -0.023 | 321.0641 | 4.41 |
| $K = 1A^- \leftarrow 2A^+$ | | | | | 12 | 438604.279 | -0.052 | 387.4393 | 6.05 | 11 | 196708.163 | -0.016 | 337.2061 | 4.37 |
| 2 | 278995.380 | -0.019 | 198.0674 | 1.57 | 13 | 436434.204 | -0.070 | 406.4820 | 6.56 | 12 | 196536.941 | 0.002 | 354.7989 | 4.23 |
| 3 | 279956.786 | -0.034 | 202.5130 | 2.71 | 14 | 434098.345 | -0.086 | 426.9868 | 7.02 | 13 | 195826.599 | 0.021 | 373.8400 | 4.02 |
| 4 | 281230.232 | -0.032 | 208.4394 | 3.69 | 15 | 431597.266 | -0.045 | 448.9529 | 7.45 | 14 | 194522.702 | 0.030 | 394.3271 | 3.76 |
| 5 | 282808.303 | 0.001 | 215.8459 | 4.56 | 16 | 428931.518 | -0.058 | 472.3797 | 7.84 | 15 | 192587.146 | 0.065 | 416.2584 | 3.48 |
| 6 | 284681.420 | -0.008 | 224.7314 | 5.35 | 17 | 426102.057 | -0.024 | 497.2664 | 8.20 | 16 | 189997.639 | 0.087 | 439.6327 | 3.19 |
| 7 | 286837.932 | 0.002 | 235.0947 | 6.03 | 18 | 423109.984 | 0.076 | 523.6121 | 8.52 | 17 | 186746.238 | 0.093* | 464.4490 | 2.91 |
| 8 | 289263.767 | 0.013 | 246.9341 | 6.62 | 19 | 419956.437 | 0.010 | 551.4160 | 8.81 | 18 | 182836.992 | 0.098* | 490.7066 | 2.65 |
| 9 | 291942.398 | 0.017 | 260.2481 | 7.11 | 20 | 416643.409 | 0.063 | 580.6771 | 9.07 | 19 | 178283.299 | 0.113 | 518.4052 | 2.41 |
| 10 | 294854.758 | 0.023 | 275.0347 | 7.48 | 21 | 413172.854 | 0.075 | 611.3945 | 9.30 | 20 | 173105.288 | 0.078 | 547.5445 | 2.19 |
| 11 | 297979.183 | 0.038 | 291.2916 | 7.73 | 22 | 409547.460 | 0.138 | 643.5669 | 9.49 | 21 | 167327.772 | 0.056 | 578.1243 | 2.00 |
| 12 | 301291.440 | 0.040 | 309.0166 | 7.87 | 23 | 405770.350 | 0.214 | 677.1933 | 9.66 | 22 | 160978.207 | 0.025 | 610.1444 | 1.84 |
| 13 | 304764.977 | 0.060 | 328.2070 | 7.89 | 24 | 401845.412 | 0.374 | 712.2725 | 9.79 | 23 | 154085.295 | -0.098 | 643.6047 | 1.71 |
| 14 | 308371.140 | 0.066 | 348.8601 | 7.80 | 25 | 397777.245 | 0.648 | 748.8031 | 9.89 | $K = 5 \leftarrow 4\ E$ | | | | |
| 15 | 312079.768 | 0.046 | 370.9729 | 7.59 | $K = 6A^- \leftarrow 5A^+$ | | | | | 5 | 359823.998 | -0.151 | 267.8567 | 1.88 |
| 16 | 315859.926 | 0.034 | 394.5425 | 7.28 | 6 | 240556.900 | -0.222 | 311.9391 | 1.79 | 6 | 361389.807 | -0.130 | 276.7070 | 3.55 |
| 17 | 319680.721 | 0.010 | 419.5656 | 6.88 | 7 | 242803.110 | -0.051 | 322.2739 | 3.36 | 7 | 363338.183 | -0.073 | 287.0368 | 5.13 |
| 18 | 323512.467 | -0.016 | 446.0390 | 6.40 | 8 | 245341.857 | -0.083 | 334.0834 | 4.69 | 8 | 365723.002 | -0.031 | 298.8480 | 6.67 |
| 19 | 327327.865 | -0.011 | 473.9597 | 5.85 | 9 | 248162.870 | -0.034 | 347.3667 | 5.87 | 9 | 368602.209 | -0.003 | 312.1425 | 8.20 |
| 20 | 331103.093 | -0.048 | 503.3246 | 5.26 | 10 | 251254.638 | 0.016 | 362.1231 | 6.92 | 10 | 372034.917 | 0.017 | 326.9224 | 9.75 |
| 21 | 334819.167 | -0.078 | 534.1307 | 4.65 | 11 | 254604.971 | 0.026 | 378.3518 | 7.86 | 11 | 376078.002 | 0.035 | 343.1893 | 11.30 |
| 22 | 338462.770 | -0.038 | 566.3753 | 4.03 | 12 | 258201.232 | 0.045 | 396.0519 | 8.71 | 12 | 380782.577 | 0.022 | 360.9447 | 12.86 |
| 23 | 342026.736 | -0.002 | 600.0558 | 3.41 | 13 | 262030.352 | 0.020 | 415.2224 | 9.48 | 13 | 386191.031 | 0.014 | 380.1899 | 14.42 |
| 24 | 345510.545 | 0.053 | 635.1700 | 2.83 | 14 | 266079.246 | -0.016 | 435.8622 | 10.18 | 14 | 392334.825 | 0.009 | 400.9254 | 15.98 |
| 25 | 348920.130 | 0.190 | 671.7157 | 2.28 | 15 | 270334.949 | -0.030 | 457.9703 | 10.82 | 15 | 399233.665 | -0.036 | 423.1514 | 17.51 |
| $K = 5A^- \leftarrow 4A^+$ | | | | | 16 | 274784.862 | 0.003 | 481.5455 | 11.40 | 16 | 406896.139 | -0.045 | 446.8676 | 19.03 |
| 5 | 449130.007 | -0.026 | 295.1203 | 1.10 | 17 | 279417.144 | 0.036* | 506.5867 | 11.95 | 17 | 415321.022 | -0.073 | 472.0734 | 20.51 |
| 6 | 448127.146 | -0.022 | 303.9150 | 2.05 | 18 | 284220.973 | 0.143* | 533.0927 | 12.45 | 18 | 424499.686 | -0.087 | 498.7676 | 21.96 |
| 7 | 446957.179 | -0.042 | 314.1749 | 2.88 | 19 | 289186.299 | -0.210 | 561.0622 | 12.93 | 19 | 434418.375 | -0.095 | 526.9490 | 23.39 |
| 8 | 445620.211 | -0.020 | 325.8996 | 3.62 | 20 | 294306.148 | -0.176 | 590.4941 | 13.38 | 20 | 445060.509 | -0.098 | 556.6160 | 24.78 |
| 9 | 444116.231 | -0.023 | 339.0889 | 4.30 | 21 | 299574.333 | -0.173 | 621.3872 | 13.82 | 21 | 456408.584 | -0.053 | 587.7670 | 26.16 |

Table 4 (*continued*)

| $J$ | $\nu_{obs}$ (MHz) | (o-c)[a] | $E_u$ (cm$^{-1}$)[b] | $\mu^2S$[c] | $J$ | $\nu_{obs}$ (MHz) | (o-c)[a] | $E_u$ (cm$^{-1}$)[b] | $\mu^2S$[c] | $J$ | $\nu_{obs}$ (MHz) | (o-c)[c] | $E_u$ (cm$^{-1}$)[b] | $\mu^2S$[c] |
|---|---|---|---|---|---|---|---|---|---|---|---|---|---|---|
| $K = 5 \leftarrow 4$ $E$, cont'd | | | | | $K = -2 \leftarrow -3$ $E$, cont'd | | | | | $K = -7 \leftarrow -6$ $E$, cont'd | | | | |
| 22 | 468445.364 | -0.079 | 620.4004 | 27.51 | 13 | 359471.092 | 0.043 | 349.2613 | 3.97 | 12 | 318253.118 | -0.133 | 442.1627 | 7.23 |
| 23 | 481155.219 | -0.027 | 654.5146 | 28.86 | 14 | 358335.117 | 0.094 | 369.7378 | 3.44 | 13 | 320716.169 | -0.122 | 461.2941 | 8.01 |
| 24 | 494524.107 | 0.039 | 690.1078 | 30.19 | 15 | 356860.223 | 0.185 | 391.6588 | 2.87 | 14 | 323209.566 | -0.149 | 481.8899 | 8.71 |
| 25 | 508539.887 | 0.063 | 727.1785 | 31.52 | 16 | 355058.135 | 0.208 | 415.0232 | 2.31 | 15 | 325706.814 | -0.111 | 503.9485 | 9.33 |
| $K = -2 \leftarrow -3$ $E$, | | | | | 17 | 352959.674 | 0.215 | 439.8306 | 1.78 | 16 | 328181.849 | -0.096 | 527.4685 | 9.89 |
| 3 | 358577.961 | 0.019 | 224.4791 | 1.58 | 18 | 350618.503 | 0.135 | 466.0815 | 1.28 | 17 | 330609.679 | -0.062 | 552.4483 | 10.39 |
| 4 | 358994.904 | 0.035 | 230.3719 | 2.73 | 19 | 348120.980 | -0.085 | 493.7775 | 0.85 | 18 | 332966.497 | -0.011 | 578.8863 | 10.84 |
| 5 | 359458.495 | 0.026 | 237.7348 | 3.61 | 20 | 345610.011 | -0.565 | 522.9221 | 0.48 | 19 | 335229.954 | 0.008 | 606.7809 | 11.24 |
| 6 | 359931.048 | -0.012 | 246.5658 | 4.28 | 21 | 343345.016 | -1.600* | 553.5230 | 0.19 | 20 | 337379.679 | 0.155 | 636.1304 | 11.60 |
| 7 | 360368.660 | -0.002 | 256.8624 | 4.75 | $K = -7 \leftarrow -6$ $E$ | | | | | 21 | 339396.805 | 0.092 | 666.9333 | 11.91 |
| 8 | 360722.317 | -0.007 | 268.6221 | 5.03 | 7 | 307325.930 | 0.081 | 368.5170 | 1.66 | 22 | 341265.312 | 0.104 | 699.1877 | 12.19 |
| 9 | 360940.086 | -0.013 | 281.8418 | 5.12 | 8 | 309239.962 | 0.022 | 380.3073 | 3.07 | 23 | 342971.165 | 0.041 | 732.8921 | 12.44 |
| 10 | 360969.692 | -0.005 | 296.5186 | 5.04 | 9 | 311314.741 | -0.039 | 393.5681 | 4.31 | 24 | 344503.061 | -0.117 | 768.0448 | 12.66 |
| 11 | 360761.723 | 0.000 | 312.6493 | 4.80 | 10 | 313525.783 | -0.068 | 408.2984 | 5.40 | 25 | 345852.502 | -0.334 | 804.6440 | 12.84 |
| 12 | 360273.241 | 0.036 | 330.2311 | 4.43 | 11 | 315847.381 | -0.108 | 424.4971 | 6.37 | | | | | |

[a] Observed minus calculated line frequencies in MHz. Asterisks indicate blended lines or lines with large residuals that were excluded from the fit.

[b] Calculated upper level energies in cm$^{-1}$. The zero reference energy is taken as the $J = K = 0$ $A$+ $v_t = 0$ level, calculated to lie 104.8883 cm$^{-1}$ above the bottom of the torsional barrier.

[c] $S$ is the transition strength. Relative line intensities are given by the product of $\mu^2S$ and the appropriate Boltzmann factor. The dipole moment components were taken as $\mu_a = 0.836$ and $\mu_b = -1.439$ Debye.

[d] The $J = 6$ to 15 $A^+ \leftarrow A^-$ components for the $K = 6 \leftarrow 5$ $A$ $v_t = 1$ $Q$-branch are not listed separately since the $K$-type doubling is not resolved and the (o-c) values for the two components differ by less than 0.1 MHz.

**Table 5**

Principal fitted molecular parameters (in $cm^{-1}$) for $^{13}CH_3OD$ and comparison to related methanol isotopologues

| Param | $^{13}CH_3OD$ | $CH_3OD$ | $CH_3OH$ | $^{13}CH_3OH$ |
|---|---|---|---|---|
| $A$ | 3.675259(77) | 3.673981(5) | 4.2537233(71) | 4.2538428(51) |
| $B$ | 0.7624739(18) | 0.7823356(4) | 0.8236523(70) | 0.8034196(51) |
| $C$ | 0.71537708(83) | 0.7328606(4) | 0.7925575(71) | 0.7737925(50) |
| $F$ | 17.41869(15) | 17.42806(1) | 27.6468464(28) | 27.64201624(70) |
| ρ[a] | 0.6959(17) | 0.6993443(2) | 0.8102062230(37) | 0.8101648121(45) |
| $V_3$ | 366.034(37) | 366.3400(1) | 373.544746(12) | 373.741301(27) |
| $D_{ab}$ | 0.0268334(83) | 0.028047(3) | −0.0038095(38) | −0.0043475(87) |

[a] Dimensionless.